\documentclass[12pt]{article}
\pdfoutput=1

\usepackage{verbatim}
\usepackage{amsmath}
\usepackage{amssymb}
\usepackage{graphicx}
\usepackage{cite}
\usepackage{subfig}
\usepackage{setspace}
\usepackage{url}
\usepackage[percent]{overpic}
\usepackage{slashed}
\usepackage{authblk}
\usepackage{footnote}
\makesavenoteenv{tabular}
\makesavenoteenv{table}

\usepackage{xspace}
\newcommand{\Fermi}[0]{\textit{Fermi}\xspace}

\usepackage{fullpage}
\usepackage{hyperref}

\def\ie{{\it i.e.}}
\def\eg{{\it e.g.}}
\def\etc{{\it etc}}
\def\etal{{\it et al.}}

\def\ba{\begin{array}}       \def\ea{\end{array}}
\def\bea{\begin{eqnarray}}   \def\eea{\end{eqnarray}}
\def\bit{\begin{itemize}}    \def\eit{\end{itemize}}
\def\beq{\begin{equation}}   \def\eeq{\end{equation}}
\def\ben{\begin{enumerate}}  \def\een{\end{enumerate}}
\def\atversim#1#2{\lower0.7ex\vbox{\baselineskip\zatskip\lineskip\zatskip
  \lineskiplimit 0pt\ialign{$\matth#1\hfil##\hfil$\crcr#2\crcr\sim\crcr}}}

\def\nn{\nonumber}

\def\k{\kappa}
\def\l{\lambda}
\def\b{\beta}

\def\Fermi{\,{\it Fermi}}

\title{Towards a Supersymmetric Description of the {\it Fermi} Galactic Center Excess}
\date{}
\author{M. Cahill-Rowley$^1$, J.S. Gainer$^2$, J.L. Hewett$^1$ and T.G. Rizzo$^1$}
\affil{$^1$ SLAC National Accelerator Laboratory, Menlo Park, CA, 94025, USA\footnote{mrowley,hewett,rizzo@slac.stanford.edu}\\
$^2$ Physics Department, University of Florida, Gainesville, FL 32611, USA\footnote{jgainer@ufl.edu}}

\begin{document}

\rightline{\vbox{\halign{&#\hfil\cr
&SLAC-PUB-16070\cr
}}}


{\let\newpage\relax\maketitle}

\begin{abstract}
We attempt to build a model that describes the {\it Fermi} galactic gamma-ray excess (FGCE) within a UV-complete 
Supersymmetric framework; we find this to be highly non-trivial. At the very least a successful Supersymmetric explanation 
must have several important ingredients in order to fit the data and satisfy other theoretical and experimental constraints. 
Under the assumption that a {\it single} annihilation mediator is responsible for both the observed relic density 
as well as the FGCE, we show that the requirements are not easily satisfied in many TeV-scale SUSY 
models, but can be met with some model building effort in the general NMSSM with $\sim 10$ parameters beyond the MSSM. We find 
that the data selects a particular region of the parameter space with a mostly singlino lightest Supersymmetric particle and a
relatively light CP-odd Higgs boson that acts as the mediator for dark matter annihilation. 
We study the predictions for various observables within this parameter 
space, and find that searches for this light CP-odd state at the LHC, as well as searches for the direct detection of dark matter, are 
likely to be quite challenging.  It is possible that a signature could be observed in the flavor sector; however,
indirect detection remains the best probe of this scenario.


\end{abstract}

\pagebreak

\section{Introduction and Background}

Observations indicate that 85\% of the matter in the universe is non-luminous and point towards the existence of cold dark matter
particles.  Determining the identify of dark matter (DM) is one of the most pressing issues before us today.  An important method of searching for dark matter is indirect detection, which searches for DM annihilating into a pair of Standard Model (SM) particles, producing an excess of gamma rays or antiparticles.  The gamma rays can appear as
a distinct line at an energy corresponding to the mass of the annihilating dark matter particles, or as a continuous energy spectrum created by radiation from scattering, hadronization, and decay of the SM final states.  Several anomalies with these features have recently appeared in 
data sets\cite{positronexcess,130line,earlygce,recentgce}.  However, it is unfortunately a fact of life that indirect
signals of dark matter can be ambiguous, and a dark matter explanation for many of the observed excesses to-date 
competes with more mundane astrophysical sources and instrumental backgrounds\cite{HooperSSI}.

The \Fermi~ Gamma-Ray Space Telescope has performed an all-sky survey of gamma rays\cite{Fermiref} and
has searched for dark matter signals from the galactic center, as well as from faint dwarf galaxies.  As data accumulates, the latter pursuit
has proven to provide relatively background-free constraints on dark matter candidates\cite{Fermidwarfs}.  
However, an excess of gamma rays from the galactic center (GC) has become increasingly intriguing as its characterization has improved \cite{earlygce,recentgce}.  
The spectrum and angular distributions of this excess are well described by annihilating dark matter, with an energy spectrum peaking at
1-3 GeV, and an approximate spherical distribution about the center of the Milky Way. The flux displays the characteristics of a NFW halo 
profile of $\rho\sim r^{-\gamma}$ with $\gamma=1.1-1.3$.  A fit to the spectrum is consistent with 30-40 GeV dark matter particles
annihilating dominantly into $b\bar b$ final states (a lighter 7-10 GeV particle annihilating to taus is also possible, but somewhat disfavored),
with a cross section of $\sigma v=1-2\times 10^{-26}$~cm$^3$s$^{-1}$.  This value of the cross section is tantalizingly consistent with 
that expected
for a Weakly Interacting Massive Particle (WIMP) thermal relic, which is a particularly interesting DM candidate since it can account for the observed DM relic density through the simple mechanism of thermal freeze-out.
Accordingly, several ideas have recently been put forward to explore a particle
physics explanation of this excess in the center of our galaxy\cite{bunch,bunchhoop,buncha,nmssmkath}.

Recently, a comprehensive study of model systematics for galactic diffuse emission has been performed \cite{calore} with the
result that an excess of gamma-rays from the GC persists in the {\it Fermi} data.  However, the fit to the dark matter hypothesis in this analysis implies
a slightly larger value of $49^{+6.4}_{-5.4}$ GeV for the mass of the dark matter candidate, with an upper limit of roughly 70 GeV at 99\% C.L.

Given the significance of the galactic center excess observed by \Fermi, and the apparent match of its characteristics to WIMP 
dark matter expectations, it is important to explore whether or not it can be accommodated by a UV-complete 
theoretical framework, such as Supersymmetry (SUSY).  SUSY with R-Parity conservation is the most appealing and widely 
examined scenario for physics beyond the SM and naturally incorporates a DM candidate, the lightest Supersymmetric particle (LSP). 
Within the framework of conventional thermal WIMP relics, the 
LSP is generally identified with the lightest neutralino, as is the case in the Minimal Supersymmetric SM (MSSM), although other possibilities 
exist in more extensive structures.  The purpose of this paper is to determine whether
the present-day pair annihilation of a neutralino LSP in the galactic center can be the source of the observed  
\Fermi~excess.  Our results show that this exercise is not trivial, and that specific requirements are placed on model building efforts which attempt to realize this possibility within the SUSY framework.  In particular, it appears that fermionic dark matter annihilating via
a relatively light pseudoscalar mediator is the best option, as recently discussed by others\cite{buncha,nmssmkath}.

As part of our model building efforts, we first examined a number of SUSY scenarios that failed to reproduce the GC excess.
The 19-parameter p(henomenological) MSSM\cite{pmssm1} does not have enough freedom to explain the GC excess, assuming a thermal history for the LSP.
A model with Dirac gauginos\cite{diracgauginos} requires t-channel exchange of a light charged
sparticle, in conflict with LEP, TRISTAN and LHC bounds.  Extending both the matter and gauge content of SUSY within a GUT
context, such as $E_6$\cite{physrep}, leads to an annihilation cross section that is far too small.
As each of these cases teaches a useful lesson, we briefly summarize our findings for each case in Section 2.

After investigating these options, we examine the Next-to-MSSM (NMSSM), which incorporates an additional gauge singlet superfield\cite{ellrev}.
This model is attractive in its own right as it naturally generates the $\mu$-term in the superpotential and easily accommodates the 
mass of the observed Higgs boson\cite{ellrev}.  The extra superfield results in an additional neutralino - the singlino - and two extra Higgs fields,
one CP-odd and one CP-even.  The extra field content allows a wider range of dark matter properties and annihilation channels\cite{nmssmdm}.
The most common form of this model incorporates a $Z_3$ symmetry that eliminates terms with
a mass dimension in the superpotential. With this $Z_3$ symmetry imposed, we find that pseudoscalar exchange alone is not sufficient to describe the observations.
A sliver of parameter space with significant annihilation through both the $Z$ boson
and light pseudoscalar $a$ channels has recently been shown to accommodate\cite{nmssmkath} 
the \Fermi~galactic center excess (FGCE). We summarize our results for the $Z_3$ option in Section 3.

We next expand our quest to one last SUSY model - the general NMSSM, which incorporates five additional parameters over the $Z_3$
case.  We find that the FGCE is well-described and selects a well-defined region of the parameter space within this model.  This region is found
to be consistent with other measurements and we study some of its predictions for future experiments.  Our results are discussed in
Section 4.

For most of our analysis we employ the results of \cite{recentgce} as discussed above with a 30-40 GeV dark matter candidate. However,
for the UV-complete scenarios that do not offer a good fit to this description, we comment on the possibility of extending the dark matter mass
range to 70 GeV (but with similar annihilation cross sections) as determined by Calore \etal \cite{calore} and show that this possibility does not
influence our results. 

In Section 5 we present our summary, where
we conclude that the most robust SUSY scenario describing the FGCE is the general NMSSM.  
Within this scenario, we find that the data selects a specific region of parameter space and that
the signatures, such as in Higgs physics at the LHC and in direct detection, are small and will be difficult, if not
impossible, to detect.  It is possible that
indirect detection thus provides the best probe of this scenario.

\section {Initial Model Building Attempts - Lessons Learned}

Here, we provide a brief overview of the set of well-motivated SUSY models that we investigated in light of the FGCE.  All of these
scenarios fail to describe the observed excess while retaining consistency with other measurements.
We believe that it is instructive to understand why each case falls short of explaining the excess and will incorporate these lessons in constructing the successful scenario presented below.

\subsection {pMSSM}

The pMSSM\cite{pmssm1} is a natural place to begin the search for a plausible SUSY scenario that can explain the FGCE since it contains the
minimal particle content, yet enjoys 
a reasonable amount of freedom in its parameter space. Here, we consider a version of
the pMSSM with a Majorana neutralino LSP ($\chi_1^0$) as part of a 19-dimensional subspace of the full $\sim $100-parameter R-Parity 
conserving MSSM.  A set of experimentally-motivated constraints have been imposed: ($a$) The soft-breaking parameters are taken to 
be real so that no new sources of 
CP violation are present, ($b$) MFV holds at the TeV scale so that flavor physics is essentially controlled by the CKM matrix, ($c$)  
the first and second generation sfermions are assumed to be degenerate for each sparticle type, and ($d$) the $A$-terms and Yukawa couplings of these two 
generations are assumed to be negligible.  In recent work\cite{uspmssm}, we have generated two sets of pMSSM models (\ie, unique points in this parameter space) with a neutralino LSP. These samples are consistent with collider, flavor, cosmological and precision measurements. We employ the ``low-FT'' model set, which has a large sample of LSPs with masses between 30 and 40 GeV, for the present study. Other than increasing the number of light LSPs, the fine-tuning requirements placed on this model set have no effect on the properties of LSPs within this mass range.

It is challenging for a $\sim 30-40$ GeV neutralino LSP  to 
generate the observed dark matter thermal relic density in the pMSSM. 
Pure winos or Higgsinos of this mass annihilate very efficiently and yield extremely small values for the relic density; 
they are also excluded by LEP data since 
the LSP will necessarily be accompanied by a corresponding nearby charged state, \ie, a chargino.
Pure binos require a t-channel exchange of a light sfermion, which is frequently in tension with LEP or LHC bounds on light charged sparticles.   
A mostly-bino LSP with a mass of $\sim 35$ GeV can annihilate through an s-channel Higgs (or H/A) boson if it contains a significant Higgsino component. However, the Higgsino component also controls the coupling to the $Z$ boson, so that $Z$ exchange is dominant in the mass region of interest (large couplings of the LSP to the Higgs are also constrained by null results from direct detection experiments). The LSP coupling to the $Z$ boson is constrained by limits on the spin-dependent dark matter direct 
detection cross section, and by measurements of the invisible width of the $Z$  ($\Delta\Gamma_{inv} \lesssim 2$ MeV). Interestingly, these combined constraints exclude a mixed bino-Higgsino LSP with a mass below $\sim 31$ GeV.

\begin{figure}[htbp]
\centerline{\includegraphics[width=5.0in]{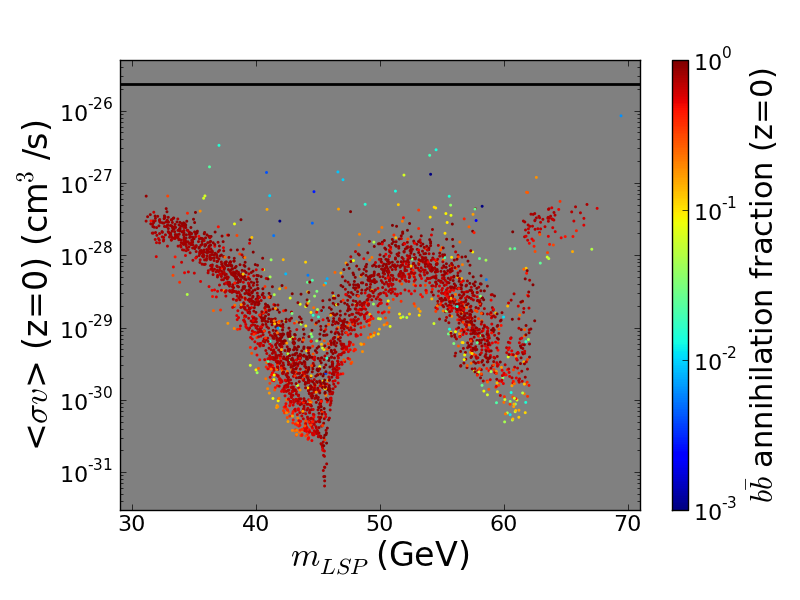}}
\centerline{\includegraphics[width=5.0in]{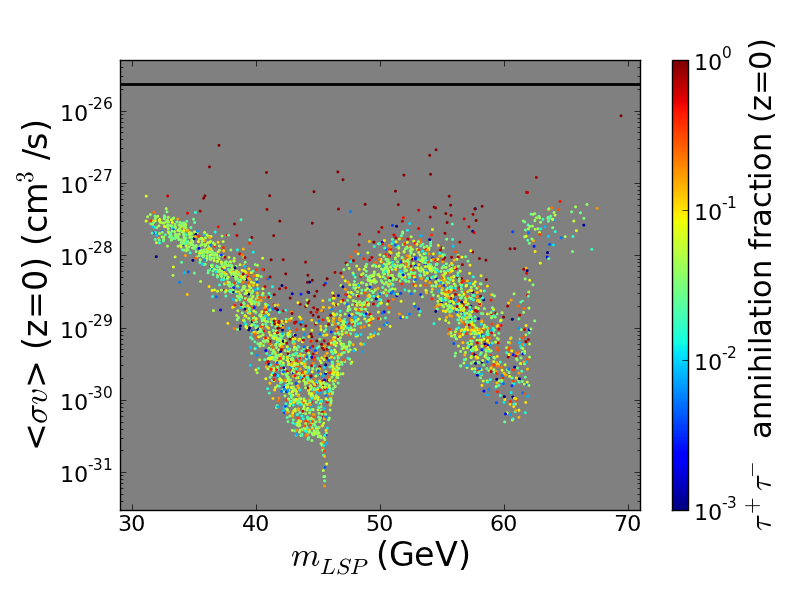}}
\caption{Present-day value of the thermally-averaged DM annihilation cross section, $<\!\sigma v\!>$, in units of cm$^3$s$^{-1}$ as a function of the 
LSP mass for models with $\Omega_{LSP} = \Omega_{DM}$ in the $b\bar b$ (top) and  $\tau^+\tau^-$ (bottom) channels.}
\label{fig1}
\end{figure}

We now examine the annihilation cross section for a LSP with a bino-Higgsino admixture.
The top (bottom) panel in Fig.~\ref{fig1} displays the present-day value of $<\!\sigma v\!>$ for the LSPs in our low-FT pMSSM model set annihilating
in the $b\bar b$ ($\tau^+\tau^-$) channel. All of these models predict the correct LSP abundance, and we have selected only models with LSP masses in the interesting range. The panels show the complete calculation as performed by DarkSUSY\cite{darksusy}. Looking first at the 30-40 GeV mass range,
we find that the $Z$ exchange contribution alone is a reasonable estimator of the total annihilation rate in most cases, and predicts an annihilation rate which is 2 orders below the value of $<\!\sigma v\!>$ during freeze-out (and therefore below the value needed to explain the FGCE). Although some points have significantly larger annihilation rates than predicted by the $Z$ exchange diagram, these models annihilate almost exclusively to $\tau^+ \tau^-$  through t-channel exchange of light staus. Bottom squarks light enough to give comparable annihilation rates are strongly excluded by collider limits. The present-day annihilation cross section decreases with increasing LSP mass because heavier LSPs are able to annihilate through the $Z$ pole in the early universe, reducing the Higgsino content required to achieve the correct relic density. Since the $Z$ pole becomes kinematically inaccessible at late times (due to low LSP velocities), the present-day annihilation cross section for heavier LSPs is suppressed by their lower Higgsino content.

If we allow the LSP mass to increase further into the $\sim 40-70$ GeV range as suggested by \cite{calore}, there are several interesting features
observed in this Figure.  As $\sim M_Z/2$ is approached from below, the Higgsino content of the LSP also  continues to decrease in order to obtain the correct relic density during freeze-out. This results in a further shrinking of 
the annihilation cross section observed today, down into and below the $\sim 10^{-30}$ cm$^{3}$ sec$^{-1}$ range. This is 
far too small to be compatible with the observed GC flux. Note that the rate slowly recovers beyond the $Z$-pole, 
exceeding $\sim 10^{-28}$ cm$^{3}$ sec$^{-1}$ for masses near $\sim 55$ GeV. Once the LSP mass appreciably exceeds 
$\sim M_Z/2$ then the cross section comes under the influence of the SM-like Higgs resonance at $\sim 125$ GeV. Since the 
Higgs essentially couples to the product of the bino and Higgsino content of the LSP, the LSP again becomes nearly pure bino 
at resonance and the annihilation cross section today once again drops appreciably. Above the Higgs pole, the annihilation rate 
again recovers. For even larger masses above the Higgs peak, co-annihilation with other light sparticles, \eg, 
$\tilde \tau$'s, becomes more important, but do not lead to $b\bar b$-rich final states. In addition to the models in 
the main band we see a number of outliers with somewhat larger cross sections, however none of these yield an appreciable rate 
into $b\bar b$. From this we learn that even for dark matter in the 40-70 GeV range, the pMSSM cannot reproduce the GC signal 
if the final state is required to be $b\bar b$.  

\begin{figure}[htbp]
\vspace*{-1.5cm}
\centerline{\includegraphics[width=4.5in,angle=90]{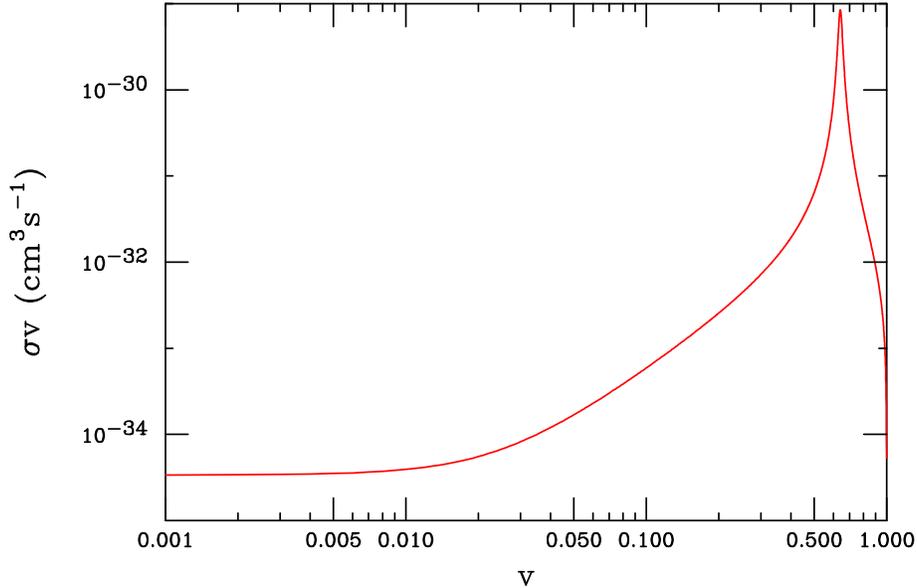}}
\vspace*{-1.50cm}
\caption{ $\sigma v$ for $\chi \chi \to b \bar b$ 
as a function of the velocity $v_\chi$, mediated by the SM $Z$ boson for $m_\chi=35$ GeV. 
Here, for purposes of demonstration, the coupling of the LSP 
to the $Z$ boson, which is expressed through its Higgsino content via the 
combination of the mixing parameters for the neutralino mass matrix $(|Z_{13}|^2-|Z_{14}|^2)$, 
is taken to be $10^{-4}$; the cross section can be appropriately 
rescaled for other values of interest.}
\label{fig1c}
\end{figure}

In Fig.~\ref{fig1c}, we show
the behavior of $\sigma v$ ({\it without thermal averaging}) for a 35 GeV Majorana LSP annihilating through $Z$ exchange, plotted as a function of the LSP velocity.  Since the 35 GeV $\chi_1^0$ annihilates through the $Z$ pole in this scenario, the cross section exhibits a strong velocity dependence.  In the early universe, the LSP velocity is  
sufficient to push the collision center-of-mass energy towards the $Z$ pole, and the annihilation cross section can take full 
advantage of the resonance to generate the observed relic density.  In the present era, the LSPs 
are moving very slowly so that their annihilation takes place away from the $Z$ pole with a cross section that is correspondingly far smaller.  LSPs with a slightly lower (higher) mass
require a somewhat larger (smaller) coupling to the $Z$, since they receive, on average, less (more) of a boost from the 
pole.  At present-day velocities, the LSPs are highly non-relativistic, and the kinematic enhancement of the annihilation rate for LSPs close the $Z$ mass (but with a mass difference larger than the $Z$ width) essentially disappears.

We have learned two lessons from this exercise: ($i$)  The FGCE cannot be explained in the pMSSM.  Given the parameter freedom
in the pMSSM, this result is likely to hold for any model in the MSSM with a Majorana neutralino LSP.  ($ii$) Since the neutralinos 
are Majorana fields and are required to annihilate through the $Z$, the cross section has a very strong velocity dependence  
due to the absence of vector couplings as well as the proximity of the $Z$ pole ($M_Z/2m_\chi \sim 1.3$).  This renders it difficult, if not
impossible, to generate the observed relic density while explaining the FGCE.
Clearly it would be advantageous to avoid this strong dependence of $<\!\sigma v\!>$ on the velocity by
building a model wherein the SM $Z$ resonance is not the dominant channel for both early universe and present-day DM annihilation.

\subsection{Dirac Gauginos}

Another possible scenario is the MSSM with Dirac gauginos, which has recently gained in popularity\cite{diracgauginos} in light of the null LHC 
SUSY search results.  In this scenario, a $\sim$ 35 GeV LSP must be almost pure bino (to the level of $\sim 10^{-4}$ or smaller) since 
any significant Higgsino component will produce a vector coupling of the LSP to the $Z$ boson.  This, in turn, yields a very large spin-independent (SI)
direct detection cross section\cite{diracdm} which is excluded by experiment.  A Higgsino admixture is thus not allowed, ensuring that 
the annihilation of the LSP does not proceed through the $Z$ boson, which avoids the problems encountered 
above in the pMSSM.  

Essentially pure Dirac binos can have a significantly larger $t-$ 
and $u-$channel induced co-annihilation cross section via sfermion exchange than in the corresponding case of a Majorana LSP, even when the
sfermion mass satisfies the LEP bounds.
The behavior of this cross section is easily understood once we recall that 
both the conventional $\mu$ and $A$-terms are absent in R-symmetric models, so that left and right sfermions do not mix. Since the 
sfermions are pure electroweak eigenstates, the cross section has two 
contributions - one from each of the $\tilde f_{L,R}$ exchanges - and the corresponding rates will be proportional to the combinations 
$N_c Y_{L,R}^4$, with $Y_{L,R}$ being the relevant hypercharges of the $\tilde f_{L,R}$. Since both sbottoms have a smaller 
hypercharge than do staus (by a factor of 3), annihilation to the $b\bar b$ final state will be suppressed by a factor of 27 relative to annihilation to $\tau^+ \tau^-$ 
when the stau and sbottom masses are equal (although we note that collider searches allow for staus to be lighter than sbottoms). 
Specifically, in the $v^2\to 0$ limit, one obtains the cross section\cite{diracdm} for the process $\chi \bar \chi \to f\bar f$:
\begin{equation}
\sigma v = {\frac{N_c ~g_1^4 m_\chi^2 ~\beta_f}{8\pi}} \Bigg ({\frac{Y_L^4}{(m_\chi^2-m_f^2+m_{\tilde f_L}^2)^2}} + L\to R \Bigg)\,,
\end{equation}
where $\beta_f^2=1-m_f^2/m_\chi^2$ and $g_1$ is the usual SM $U(1)$ hypercharge coupling. This expression shows the expected behavior. Despite the hypercharge suppression, sbottom exchange (and therefore the $b\bar{b}$ final state) can dominate if sbottoms are much lighter than the other sfermions. The problem now becomes one of rate since collider searches\cite{pdg} constrain sbottoms to be somewhat heavy, particularly for the light LSP masses required to explain the FGCE, with a limit of $m(\tilde{b}_1) \geq  300-400$ GeV being extremely conservative. Taking $m_\chi \simeq 35$ GeV and both sbottoms degenerate with masses at their smallest possible value of 300 GeV, we obtain an annihilation cross section that is too small, by a factor 
of almost $10^3$, to explain the FGCE or to obtain the observed relic density. (Note that the annihilation 
cross section during freeze-out is $\sim 10-15\%$ larger than its present-day value due to the O($v^2$) terms that were neglected in the 
above expression.) Explicitly, in the $v\to 0$ limit we obtain the result: $\sigma v \simeq 3.7 \times 10^{-29}$~cm$^3$s$^{-1}$. The annihilation rate would decrease further if heavier, more realistic sbottom masses were employed. Clearly this scenario cannot be responsible for the galactic center excess.

From the discussion above it is clear what would happen if the dark matter mass range were to be extended upward to $\sim 70$ GeV. 
Eq.~(1) shows that if we hold all other parameters fixed, then a doubling of the dark matter mass would lead to a simple four-fold 
increase in the annihilation cross section which yields an adequate rate for the $\tilde \tau$-pair final state.  
However this relatively small increase is much too meager to achieve the required annihilation into $b$-quarks, remaining  well over a factor of $\sim 100$ below the determined value. 

\subsection{{$SO(10)/SU(5)$} Singlets in {$E_6$} GUTS}
 
Another approach is to consider a SUSY Grand Unified Theory (GUT),
extending both the gauge group and matter content of the MSSM.  In this case, one of the new weakly 
interacting,  SM-singlet neutral fermions can be the LSP and potentially a DM candidate. One attractive possibility is the SUSY version of $E_6$, which can lead to an 
additional $U(1)$ gauge group at the TeV scale\cite{physrep}.  Here, the GUT-scale breaking pattern can take the form of
$E_6\to SO(10)\times U(1)_\psi$, followed by $SO(10) \to SU(5)\times U(1)_\chi$ and then, as usual, $SU(5)$ breaks to the SM. 
A linear combination of the two additional $U(1)$ groups,  
$U(1)_\theta=U(1)_\psi\cos\theta - U(1)_\chi\sin\theta$, survives to lower (\ie, TeV-scale) energies. In addition to the extended gauge 
group, the standard MSSM matter and Higgs content is also enlarged as both are contained in the fundamental {\bf 27} (and possibly 
{$\bf \overline{27}$}) representation of $E_6$. The {\bf 27} decomposes as {\bf 16}+{\bf 10}+{\bf 1} under $SO(10)$ with the SM fermions and a RH-neutrino 
composing the usual {\bf 16}. Within this scenario, one of the new light neutral SM singlet fields provides an appealing candidate for the LSP;
a particular realization of this is given by the choice of the $SO(10)$-singlet field, $S$. Since the LSP is a SM singlet, it must annihilate through the new $Z'$ gauge boson. The mass of the $Z'$ is required to be quite heavy, $M_{Z'} \gtrsim 2$ TeV, by null results from LHC resonance searches \cite{lhczprime}. Although the heavy $Z'$ mass avoids issues associated with direct detection, the invisible width of the $Z$, or the significant velocity dependence observed above in the case of the pMSSM, we will see that it also leads to a problematic suppression of the annihilation rate. Another drawback of annihilation through the $Z'$ is that $b\bar b$ is {\it not} necessarily the dominant final state for LSP annihilations.
Depending on model details, the LSP may be either a Majorana or Dirac fermion. Once this choice is made, the model contains only two free parameters, the $Z'$ mass and the value of the mixing angle $\theta$, and is therefore very predictive. Once these parameters are known, the freeze-out and present-day annihilation cross sections are easily determined.

\begin{figure}[htbp]
\centerline{\includegraphics[width=5.0in,angle=90]{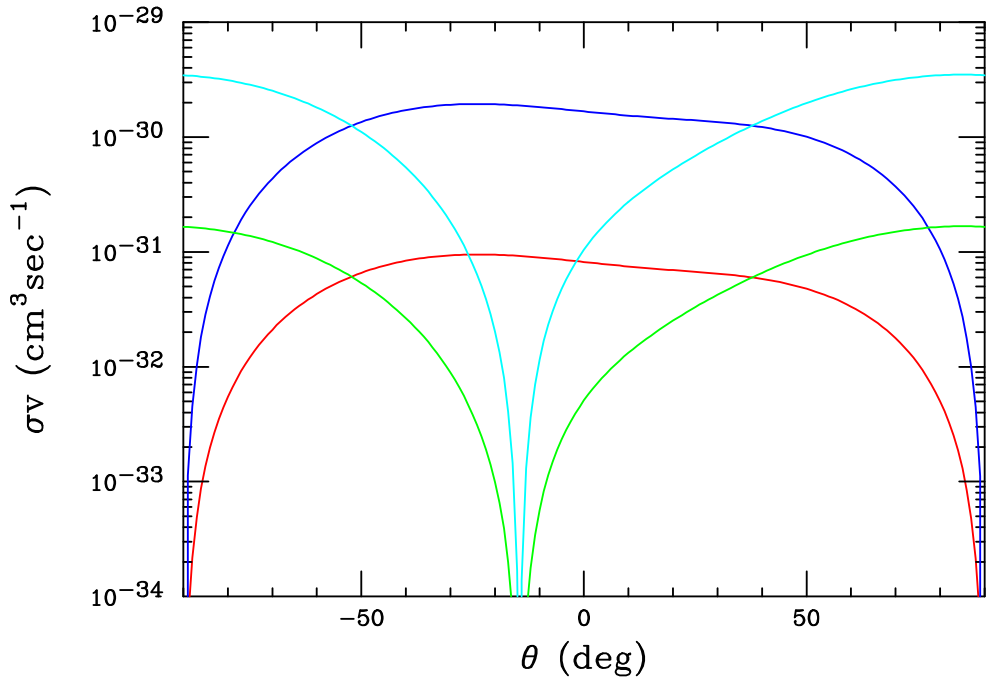}}
\vspace*{-0.90cm}
\caption{Thermal relic cross section for a 35 GeV LSP in the $E_6$ model for a $Z'$ mass of 2 TeV as a function of the mixing angle $\theta$. 
The red (blue) curve is the result obtained for the $SO(10)$ singlet field $S$ when it is Majorana (Dirac), while the green (cyan) curve represents the 
corresponding result for the $SU(5)$ singlet within the $SO(10)$ {\bf 16} representation.}  
\label{fig2}
\end{figure}

Figure~\ref{fig2} shows the thermal relic cross section to $O(v^2)$ as a function of the mixing angle $\theta$ for 35 GeV Majorana or Dirac $S$ LSPs. 
For our numerical results, we have assumed that $x=m_\chi/T_f=25$, with $T_f$ being the freeze-out temperature, 
and have taken $M_{Z'}=2$ TeV, saturating the lower limits from the LHC. Note that in this case the expansion of the thermally-averaged cross section in powers of velocity is fully valid as the annihilation takes place far away from the $Z'$ pole. We see that the cross section vanishes in the limit when the 
$Z'$ is purely a $Z'_\chi$ state as $S$ is a $SO(10)$ singlet.  Whether the LSP is Dirac or Majorana, we see that the resulting cross section 
is far too small (by at least several orders of magnitude) to predict the correct relic abundance or explain the FGCE. This results from the 
rather small gauge couplings in this class of models and from the requirement of a heavy $Z'$ paired with a light LSP, since 
the cross section roughly scales as $\sim m_\chi^2/M_{Z'}^4$.  Although the annihilation rate is much larger for a Dirac LSP than for a Majorana LSP, both results are far too small to produce the observed relic abundance or explain the FGCE. We obtain a qualitatively similar result in the case were the LSP transforms instead as the $SU(5)$ singlet within the $SO(10)$ {\bf 16} representation, although the detailed dependence of the cross 
section on $\theta$ is quite different and the cross section is now seen to vanish near $\theta \simeq -14.48^o$ as can be observed in Fig.~\ref{fig2}. Once again, the Dirac LSP case leads to a much larger cross section but also falls far short of what is required.

Increasing the dark matter mass by a factor of $\sim 2$ into the $\sim 70$ GeV range clearly does not modify our results. Given the cross section 
scaling discussed above this would yield only an additional factor of $\sim 4$ in the GC flux, thus remaining too small by roughly two 
orders of magnitude.

\section{The NMSSM}

In the  Next-to-Minimal Supersymmetric Standard Model (NMSSM),
the Higgs sector is extended by an additional gauge singlet superfield, $\hat S$.  The scalar component of this superfield obtains a vev, $s$, 
and subsequently generates the usual MSSM $\mu$ parameter dynamically through a term $\lambda \hat H_u \hat H_d \hat S$ in the superpotential, \ie, 
$\mu_{eff}=\lambda s$. The presence of the field $\hat S$ also implies the existence of an additional neutralino (the singlino), as well as one 
additional pair of weak isosinglet, CP-odd and CP-even Higgs bosons that mix with their usual MSSM counterparts.  This results in two
physical pseudoscalar bosons, $A_1=a$ and $A_2=A$ and three physical neutral CP-even Higgs bosons, $H_1=h$, $H_2$, and $H_3$, where $a$ and $h$ are the lightest
CP-odd and CP-even states, respectively.  Here we assume that $h \sim 125$ GeV is the SM-like Higgs boson discovered at the LHC.  
We identify 
the $\sim 35$ GeV LSP as being principally composed of the singlino, with small mixings with the other gauginos. The LSP can annihilate to heavy fermions (mostly $b\bar b$) by exchanging a relatively light pseudoscalar $a$, which couples to SM fields through its mixing with the MSSM pseudoscalar $A$. Reasonably large values for $\tan \beta$ are generally required to provide a large enough $ab\bar{b}$ coupling. 
In this discussion, we assume that s-channel $a$ exchange is solely responsible for producing {\it both} the observed relic density and 
the FGCE, although this need not be the case\cite{nmssmkath}. With this premise, the LSP coupling 
to the SM $Z$ boson must be extremely weak in order to avoid $Z$-mediated LSP annihilation; this requires the Higgsino content of the LSP 
to be quite small. In principle, such a `simple' scenario appears to avoid\cite{buncha} all of the pitfalls we have discussed 
in the previous section, since the LSP pair annihilation is no longer velocity suppressed, while the SI direct detection cross section has a 
very strong velocity suppression\cite{bunchhoop}. The flux from the galactic center appears to require, if anything, a present-day annihilation rate which is slightly smaller than the early universe freeze-out value, implying that $a$ must be heavier than $2m_\chi \simeq 70$ GeV so that the average center-of-mass energy of the annihilation is closer to the $a$ pole at the larger DM velocities present in the early universe. This assertion further implies that the process $h\to aa$ (where as noted above $h$ is the $\simeq 125$ GeV SM-like Higgs boson discovered at the LHC) 
cannot occur on-shell. Since the difference between the early universe and present-day annihilation rates is small, $m_a$ must
be large enough so that early universe annihilations (with $v^2 \simeq 0.10-0.15$) don't benefit too much from the $a$ pole enhancement after thermal averaging. On the other hand, if $m_a$ becomes too large, the annihilation cross sections will be too small to account for the observed relic density or the FGCE, indicating that a balance between these constraints is necessary. As we will see below, satisfying all of these conditions simultaneously is quite challenging even in the NMSSM.

\subsection{The NMSSM with {$Z_3$} Symmetry}

The simplest and most frequently studied version of the NMSSM includes a $Z_3$ symmetry that eliminates all terms with mass dimension from the superpotential.  With $\mu_{eff}$ generated as described above, the only other term\cite{ellrev} in the superpotential 
not found in the MSSM is ${\frac{\kappa}{3}} {\hat S}^3$. As a result, only a few parameters are required beyond those found in the MSSM:
$\lambda, ~\kappa, ~s$ and $A_{\lambda,\kappa}$, with the latter being the corresponding soft breaking A-terms, \ie,   
\begin{equation}
W_\mathrm{Higgs} = \lambda \widehat{S}\,\widehat{H}_u \cdot\widehat{H}_d +\frac{\kappa}{3} \widehat{S}^3\,,
\end{equation}
and, similarly,  
\begin{equation}
-{\Delta \cal L}_\mathrm{soft} = \lambda A_\lambda\, H_u \cdot H_d\; S + \frac{1}{3} \kappa A_\kappa\,S^3 + \mathrm{h.c.}\,.
\end{equation}
Interestingly, it is easy to argue that even at tree level these few parameters alone do not allow enough flexibility to construct a successful 
model along the lines discussed above where only the light pseudoscalar $a$ participates significantly in DM annihilation.
This can be most easily seen by examining the tree-level neutralino mass matrix in the $Z_3$-invariant case as is given in Eq.(2.32) of Ref.~\cite{ellrev} which we repeat here for clarity:  
\beq
{\cal M} =
\left( \ba{ccccc}
M_1 & 0 & -\frac{g_1 v_d}{\sqrt{2}} & \frac{g_1 v_u}{\sqrt{2}} & 0 \\
& M_2 & \frac{g_2 v_d}{\sqrt{2}} & -\frac{g_2 v_u}{\sqrt{2}} & 0 \\
& & 0 & -\mu_\mathrm{eff} & -\l v_u \\
& & & 0 & -\l v_d \\
& & & & 2 \k s + \mu'
\ea \right) \,.
\eeq
Here, $v_u (v_d)$ is the vev of the Higgs doublet giving mass to the up (down) sector, $\mu'=0$ in the $Z_3$ invariant case, and 
$g_{1,2}$ are the usual SM gauge couplings. Recall that the effective $\mu$ parameter is given by $\mu_{eff}=\lambda s$. 

There are several constraints that need to be imposed on this neutralino mass matrix. First, the mixing of the mostly 
singlino LSP with the Higgsino must be highly restricted, otherwise the LSP will have significant couplings to the $Z$ boson (violating our assumption that only $a$ is responsible for 
DM annihilation) and all of the problems present in the pMSSM will be encountered.
Using the results in Fig.~\ref{fig1}, we can easily determine the present-day value ($v \to 0$) of the cross section for LSP annihilation through the SM $Z$ boson. Specifically, we obtain the result $\sigma v\simeq 3.39 \times 10^{-32}$~cm$^3$s$^{-1}$ when the combination of mixing parameters 
for the neutralino mass matrix $(|Z_{13}|^2-|Z_{14}|^2)^2=10^{-5}$ and 
$m_\chi=35$ GeV are assumed, where $Z_{ij}$ is the matrix that diagonalizes Eq. (4).
(Of course, for other possible Higgsino admixtures these numerical results are simply rescaled.) Similarly, employing 
$x=m_\chi/T \simeq 20(25)$ in order to calculate the thermally-averaged cross section in the early universe, we obtain the significantly larger result of 
$\sigma v\simeq 6.56(4.67) \times 10^{-30}$~cm$^3$s$^{-1}$ under the same assumptions since the $Z$ pole is partially probed by the thermal LSP energy 
distribution. Here we see explicitly that the cross section in the early universe is significantly larger than it is today by a factor of over 100. 
We thus require that $a$ exchange remains dominant over $Z$ exchange at all times, and impose the constraint that 
$(|Z_{13}|^2-|Z_{14}|^2)^2\leq 10^{-(3-4)}$. Second, the Higgsino mass, given by $\mu_{eff}=\lambda s$, must be $>100$ GeV due to LEP 
constraints on the mass of the corresponding charged state, while the parameters $\kappa, \lambda$ must be less than $\simeq 0.7$ to satisfy perturbativity 
requirements. Finally, the LSP mass in this scenario is roughly given by $m_{LSP} \simeq 2\kappa s \simeq 35$ GeV, constraining the value of $\kappa$. 

\begin{figure}[htbp]
\centerline{\includegraphics[width=5.0in,angle=90]{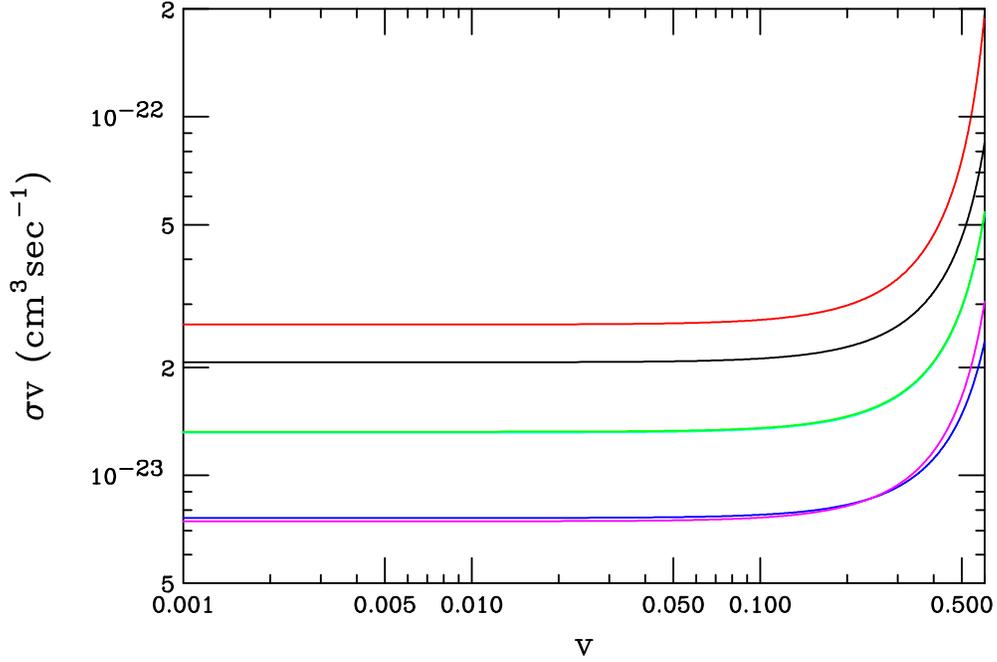}}
\vspace*{-0.90cm}
\caption{Scaled ($Q$=1) annihilation cross section for $\chi \chi \to a \to b\bar b$ as a function of $v_\chi$ assuming $m_\chi=35$ GeV for several 
different input parameter choices:  The green ($\tan \beta=40, m_a=110$ GeV, $\Gamma_a/m_a=0.02$) curve provides the reference,
with the corresponding changes in the parameters represented by the
red ($m_a=100$ GeV), blue ($m_a=120$ GeV), cyan ($\Gamma_a/m_a=0.05$), magenta ($\tan \beta=30$), and black ($\tan \beta=50$) curves.
Note that the green and cyan curves are essentially identical.}
\label{fig3}
\end{figure}

At this point, we note that the DM annihilation rate through the $a$ boson is essentially set by the combination of couplings 
$\sqrt 2 \kappa \sin \theta_a \cos \theta_a Z_{15}^2$, where $\theta_a$ measures the weak doublet content (\ie, the MSSM $A$ content) of $a$ and $Z_{15}^2 \simeq 1$ measures the singlino content of the LSP.  This implies that the 
value of $\kappa$ cannot be too small, otherwise the calculated relic density will be too large. We can see this explicitly by examining 
the annihilation cross section.  Before any thermal averaging, we find that $\sigma v$ for 
the process $\chi \chi \to a \to b\bar b$ is given by the expression 
\begin{equation}
\sigma v = {\frac{3G_F^2M_W^2m_b^2t_\beta^2}{\pi}} ~{\frac{s\beta_b (\lambda_b\lambda_\chi)^2}{(s-m_a^2)^2+ m_a^2\Gamma_a^2}} \,
\end{equation}
where $\beta_b^2=1-4m_b^2/s$, $t_\beta=\tan \beta$, $\lambda_b=\sin \theta_a$, with $\theta_a$ being the CP-odd mixing angle as described above, 
and $\lambda_\chi \simeq {\sqrt {2} \kappa \cos \theta_a} Z_{15}^2/g$ controlling the $a\chi \chi$ coupling\cite{ellrev}. To get an 
idea of the size of this cross section let us assume that the product of couplings $Q=\lambda_b\lambda_\chi=1$ and rescale the result later as 
needed;  the resulting cross section is shown in Fig.~\ref{fig3} as a function of $v_\chi$ for several different input parameter choices, as labeled.

From this Figure we learn several crucial lessons: ($i$) the cross section has a rather simple sensitivity to both $\tan \beta$ and $m_a$, 
($ii$) a lower (higher) value of $m_a$ increases (decreases) the sensitivity to the $a$ resonance at large values of $v_\chi$, and ($iii$) the  
result is not sensitive to the precise value of $\Gamma_a/m_a$ as long as this value is small and the center-of-mass energy of the annihilation
is not too close to the resonance.
To be specific for this scenario, we choose $m_a=120$ GeV, $\tan\beta=40$, and $\Gamma_a/m_a=0.01$, although the 
latter choice has very little impact as seen above.   These 
parameter choices yield a present-day thermally-averaged cross section, $<\!\sigma v\!>= 7.56\times 10^{-24}$~cm$^3$s$^{-1}$, with an early universe freeze-out value of $<\!\sigma v\!>= 9.05(8.72)\times 10^{-24}$~cm$^3$s$^{-1}$ with $x=m_\chi/T=20(25)$, assuming that $Q=1$. This result strongly constrains the value of $Q$, requiring $Q \sim 1/(300-400)$. For values of $m_a \simeq 120$ GeV we learn from \cite{buncha} that $\sin \theta_a \lesssim 0.1$ is likely 
necessary to satisfy experimental constraints when $m_A \sim 1$ TeV; assuming that $\sin \theta_a \sim 0.1$ we see that $\kappa$ cannot be too small, with values of
$\kappa \sim 1/4-1/3$ preferred, if $Q$ is to be large enough to explain both the FGCE and the observed relic density. 

\begin{figure}[htbp]
\centerline{\includegraphics[width=3.5in,angle=90]{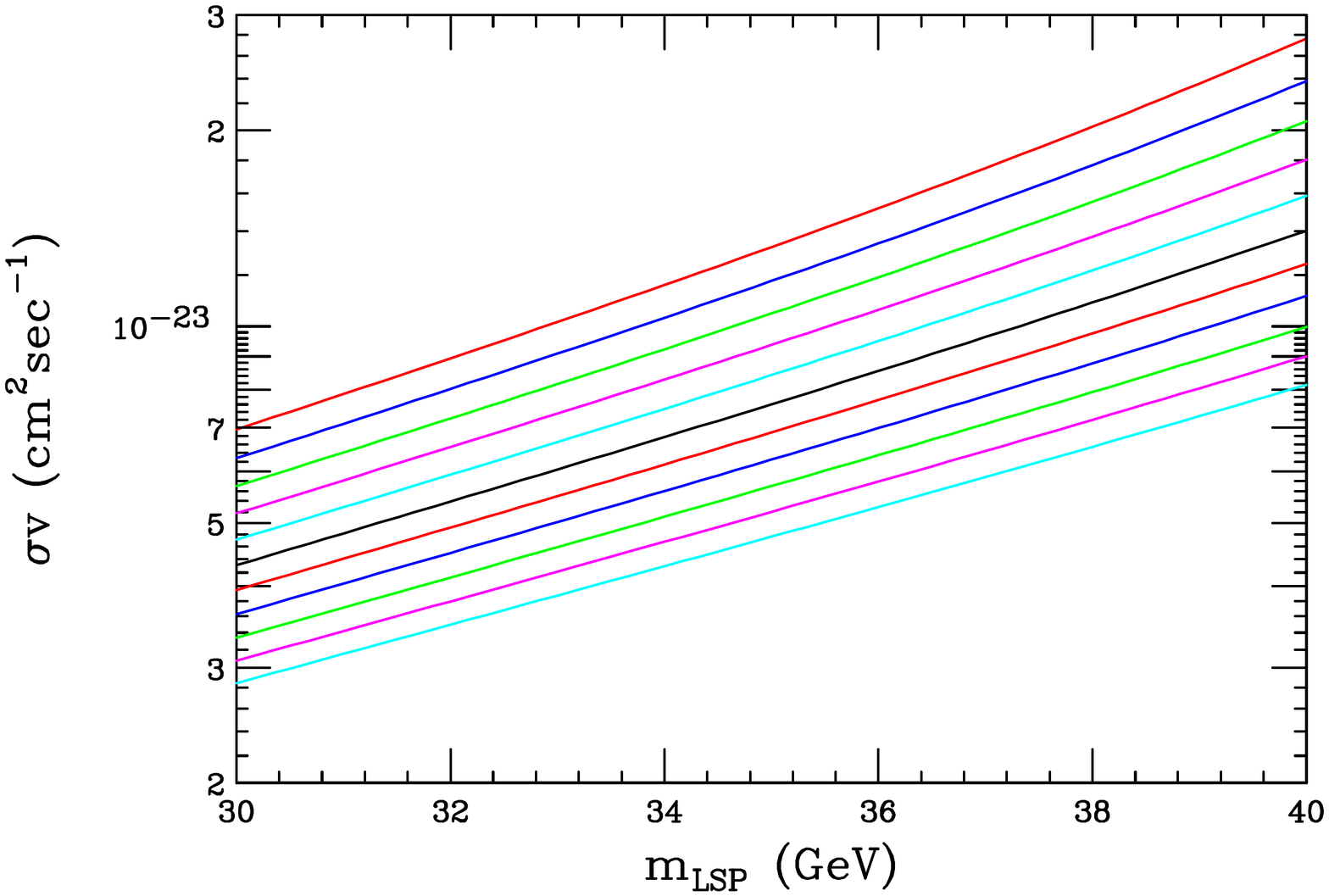}
\hspace {-1.2cm}
\includegraphics[width=3.5in,angle=90]{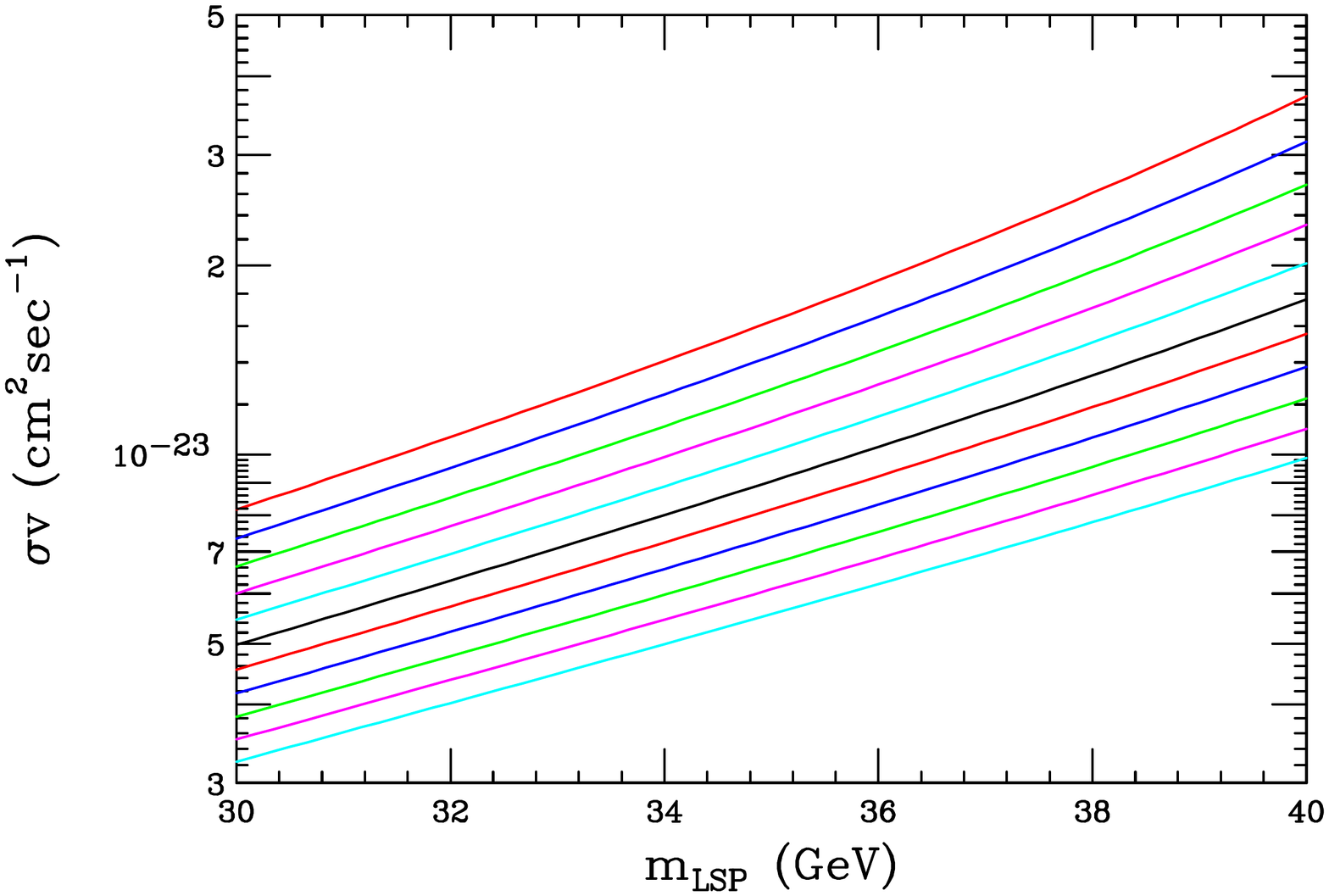}}
\vspace*{-1.0cm}
\centerline{\includegraphics[width=3.5in,angle=90]{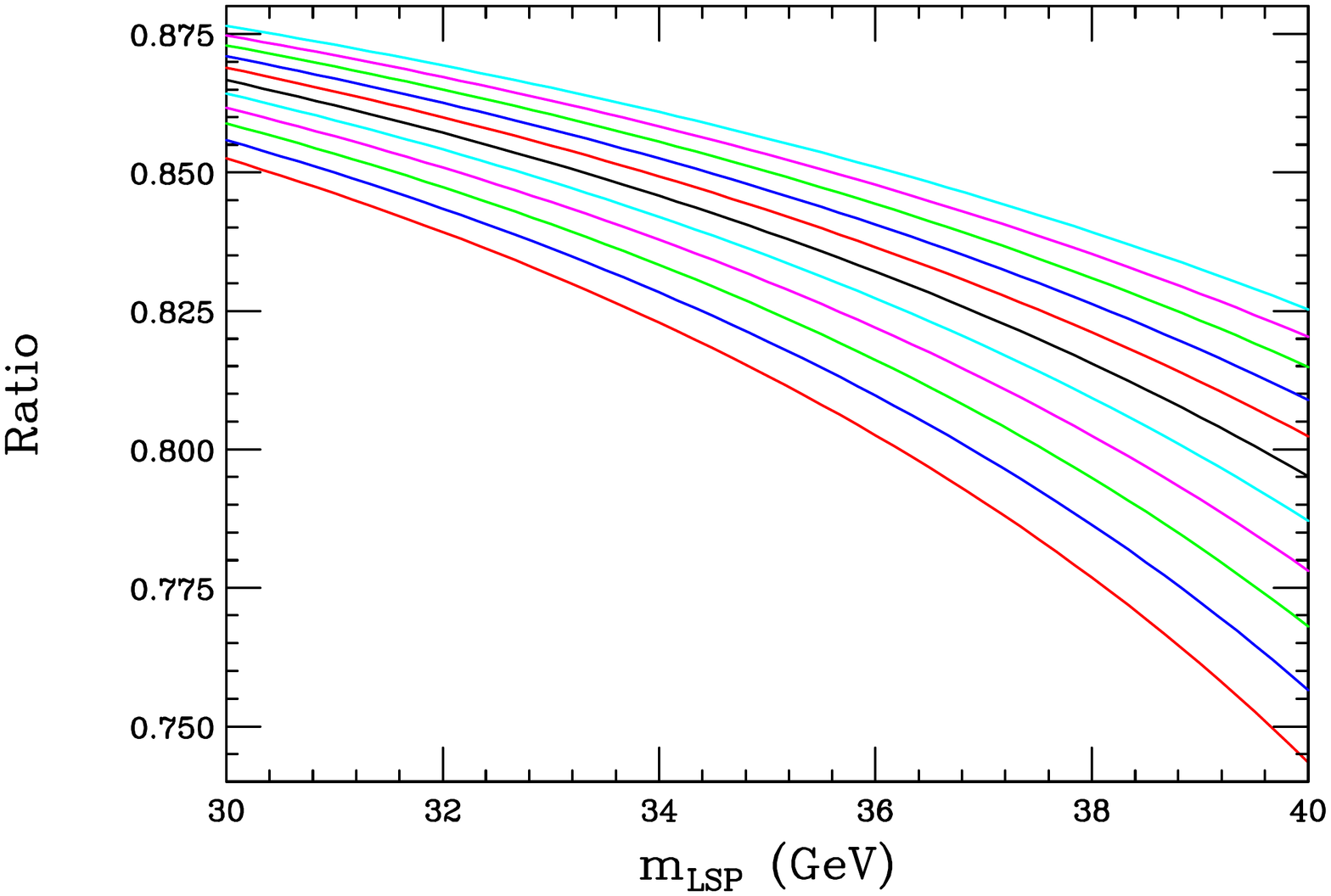}}
\vspace*{-1.00cm}
\caption{The scaled $\chi \chi \to b\bar b$ annihilation cross section as a function of $m_\chi$ computed with $Q=1$ for present-day annihilations (top left), 
for freeze-out assuming $x=m_\chi/T=20$ (top right), and the ratio of these annihilation cross sections (bottom), which is $Q$ independent. 
In the top panels, from top to bottom $m_a$ ranges from 
110 to 130 GeV in steps of 2 GeV. This mass ordering is reversed in the bottom panel.}
\label{fig4}
\end{figure}

It is instructive to examine the sensitivity of both the present-day and the early universe thermally-averaged cross sections to variations
in the values of both $m_\chi$ and $m_a$ in the region of interest; this exercise will yield information about the possible values for the remaining model parameters, such as $\kappa$. The results of these calculations are shown in Fig.~\ref{fig4}, again
taking the factor $Q$ to be unity. Here we see that these 
cross sections are reasonably sensitive to both of these mass parameters, varying by approximately an order of magnitude over the ranges of interest. 
The $Q$-independent ratio of the present-day thermally-averaged cross section to its early universe value is somewhat less sensitive to these parameters, 
but displays interesting mass dependencies. This ratio is less than unity by construction since $m_a > 2m_\chi$ as discussed above. For fixed 
$m_a$, annihilating LSPs of increasing mass start to probe the kinematic region near the $a$ pole with their thermal energy 
distribution, decreasing the ratio of the present-day annihilation cross section to its early universe value. For fixed $m_\chi$, increasing $m_a$ decreases the sensitivity to the $a$ pole region and the ratio tends towards unity. For most of the parameter values this ratio is seen to be roughly $\sim 0.8$ with a spread of approximately $\sim 10\%$, which agrees with the FGCE observations.

Although we can avoid a troublesome suppression of the present-day annihilation rate, several model building requirements combine to restrict the parameter space in this scenario. We start by considering the neutralino mixing matrix more carefully. As mentioned above, we want the LSP to be mostly singlino in order to have a large coupling to the pseudoscalar singlet. In this case, the bino and wino components of the LSP are typically small since bino-singlino and wino-singlino mixing terms are absent from the neutralino mass matrix, and because LEP results require the charged wino to be heavier than 100 GeV. We therefore assume that the bino and wino are irrelevant and consider only the Higgsino-singlino sector, which is completely specified by 4 parameters ($s, ~\kappa, ~\lambda$ and $\tan \beta$). Recalling that the LSP must have a small Higgsino component to suppress the $Z\chi\chi$ coupling, we see that for large values of $\tan \beta$, $\lambda$ must be small enough that the off-diagonal entries are much smaller than the diagonal entries, $\mu_{eff} = \lambda s$ and $2\kappa s \simeq 35$ GeV. Specifically, the mixing is roughly set by the ratio $8\lambda^2 v_u v_d/(\lambda s+2\kappa s)^2$. LEP limits require $\mu_{eff} \geq 100$ GeV, which demands a large value of s in the (necessary) case of small $\lambda$. Since the LSP mass is set by $2\kappa s \simeq m_{LSP} \simeq 35$ GeV, large values of $s$ demand small values of $|\kappa|$ to produce the correct LSP mass. This is in tension with the the requirement that $|\kappa|$ should be large enough to predict the correct annihilation cross section.  This tension indicates the failure of this approach.  While it is likely that
this complete argument can be demonstrated algebraically, we have instead performed large numerical scans (totaling $\sim 10^{10}$ points in the space) over the 4 parameters relevant for singlino-Higgsino mixing in order to verify our conclusion. No successful solutions were obtained, and we therefore conclude that the NMSSM with $Z_3$ symmetry where {\it only} the $a$ exchange is responsible for both the DM relic density and the FGCE also fails to reproduce the observations. 

We now examine how these conclusions are modified if the dark matter mass falls into the heavier $\sim 40-70$ GeV range as discussed by Calore 
\etal \cite{calore}. There are now distinct regions of parameter space to be examined, depending on the relative values of $m_a$ and 
$m_\chi$.  For the case when $m_a<2m_\chi$ the LSP pair can annihilate into the $2a$ final state.  This will dominate over the $b\bar b$ mode as 
there is no longer any helicity suppression. This subregion is thus excluded from further consideration. If $m_\chi < m_a < 2m_\chi$, 
the center of mass energy in the LSP pair annihilation process will be larger at freeze-out than it is today and 
it will occur further away from the $a$-pole and thus we expect  the LSP annihilation rate today ($z=0$) to be {\it larger} than 
in the early universe. For $m_\chi$ in the $\sim 40-70$ GeV mass range, a numerical study shows that the ratio 
$r=<\sigma v>_{FO}/<\sigma v>_{z=0}$ always lies below $\simeq 0.85-0.90$ when $m_\chi < m_a < 2m_\chi$. This conflicts with 
the requirements on the flux for the galactic center excess signal discussed above, that the current LSP annihilation rate 
should be at or below the value at freeze-out. From this discussion we then  conclude that we must 
have $m_a >2 m_\chi$, even if the LSP lies in the mass range $m_\chi = 40-70$ GeV. This being the case,  all of the arguments delineated above for LSP's in the 30-40 GeV range will still be applicable; for example, 
a very small Higgsino content of the LSP must still be maintained. In fact, for some LSP masses in this range, the constraint on 
the LSP Higgsino content would need to be further {\it tightened} due to the rather close proximity of the $Z$ pole in the annihilation process. 

Given a fixed (and large) $\tan \beta$, the annihilation cross section scales roughly as $\sim (\kappa m_\chi)^2/m_a^4$, the Higgsino content of the LSP is controlled by the ratio $8\lambda^2 v_uv_d/(\lambda s +2\kappa s)^2$, and the LSP mass is 
arises from $\simeq 2\kappa s$. It is clear how these quantities are modified when we consider LSP masses in the 
$\sim 40-70$ GeV range, and, at the very least, we don't want to further increase the tension among the parameters 
encountered above when we assumed $m_\chi \sim 35$ GeV. For example, if the LSP mass is 70 
GeV, doubling the values of both $\kappa$ and $m_a$  while leaving the corresponding values of both $\lambda$ and $s$ 
unchanged, we recover the same annihilation rate and Higgsino content that we found in the case of 
$m_\chi=35$ GeV and so the already unacceptable tension among the parameters discussed above is not increased (or improved). However, we see 
that changing the values of $\lambda$ and $s$ would only increase this tension further. Thus we conclude that for LSP 
masses in the $\sim 40-70$ GeV mass range, the $Z_3$ invariant version of the NMSSM will remain as finely-tuned and disfavored as it was 
for lighter LSP masses.

\subsection{The General NMSSM}

To go further we drop the requirement of the $Z_3$ symmetry on the NMSSM superpotential, providing greater parameter freedom. 
Although the scale-invariant feature of the $Z_3$ model is very attractive, it also poses potential domain wall problems\cite{maniatis} that
we now avoid.  Additionally, some of the attractive features of the $Z_3$ model can be recovered or improved upon in the general NMSSM with model-building effort~\cite{nmssm-zN}.  In order to most easily discern whether this more general scenario can provide a successful description of the FGCE, we first present a (mostly) tree-level discussion, which is important for providing the required transparency given the relative complexity of the parameter space.  A more detailed numerical 
analysis including all corrections and a full scan over the parameter space follows in the next section.
 
The number of additional parameters is significantly increased in the full NMSSM: First, the superpotential is augmented 
by terms which are either linear or quadratic in the singlet superfield $\hat S$. Second, corresponding additional terms are necessarily generated 
in the soft-breaking Lagrangian. Thus we find, instead of the compact expressions above, the augmented results\cite{ellrev} 
(ignoring a possible bare $\mu$ term) 
\beq
W_\mathrm{Higgs} = \lambda \widehat{S}\,\widehat{H}_u \cdot \widehat{H}_d +\xi_F \widehat{S} +\frac{1}{2} \mu^\prime \widehat{S}^2 +\frac{\kappa}{3} 
\widehat{S}^3\,,
\eeq
and 
\beq
-{\Delta \cal L}_\mathrm{soft} = \lambda A_\lambda\, H_u \cdot H_d\; S + \frac{1}{3} \kappa A_\kappa\,S^3 + m_3^2\, H_u \cdot H_d + 
\frac{1}{2}m_{S}'^2\, S^2 + \xi_S\, S+ \mathrm{h.c.}\,
\eeq
From these expressions we see that the general NMSSM scenario introduces 5 additional new parameters ($\mu^\prime, \xi_F, \xi_S, m_3, m'_S$)
beyond those encountered 
in the $Z_3$ model; only a subset of these additional parameters is required (at least at tree level) to describe the FGCE. In the brief discussion 
that follows, we assume for simplicity that $m_3^2=\xi_F=0$ in order to derive a workable model, but this need not be the case; viable models can 
be found when these parameters are non-zero and we include them below in our full scan over the parameter space.
As was shown above, the neutralino mass matrix depends on only one of these 
new parameters, $\mu^\prime$, which is sufficient to resolve the difficulties we found in the case of the $Z_3$-invariant NMSSM. 
To see this, we again assume that $M_{1,2}$ are large so that the bino and wino are decoupled, and similarly assume that the value of $\mu_{eff}$ is large enough to suppress Higgsino-singlino mixing while the value of $\lambda$ is somewhat small. This implies that large values of $s$ are necessary, meaning that $2\kappa s$ is also relatively large if $\kappa$ is not too small as previously discussed. We can, however, choose $\mu^\prime$ to have a similar magnitude to $2\kappa s$ with the opposite sign, so that these terms cancel to produced a small remainder, \ie, we assume the combination $2\kappa s +\mu^\prime \simeq 35$ GeV. Note that a fine-tuning at the few percent level is thus required to obtain the desired LSP mass.  We then perform a simple (flat) scan over parameters relevant for singlino-Higgsino mixing, employing the scan ranges $0\leq \lambda \leq 0.6$, $30 \leq \tan \beta \leq 50$, $1 \leq s \leq 10$ TeV with $0\leq \kappa/g_2 \leq 2$ and imposing the requirements discussed above. We obtain multiple valid solutions, demonstrating the viability of this approach.

Given values for the parameters ${\lambda,\kappa, \tan \beta, s, \mu^\prime}$ resulting from our scan that yield a neutralino mass matrix with the desired properties, 
we next consider the effective $2\times 2$ CP-odd scalar mass matrix at tree level. We use the convenient basis where the conventional neutral Goldstone 
boson has been removed so that the elements of this matrix are given by\cite{ellrev} 
\bea
{\cal M}_{P,11}^2 & = & \frac{2 (\mu_\mathrm{eff}\, B_\mathrm{eff} +
\widehat{m}_3^2)}{\sin 2\b}\; , \nn\\
{\cal M}_{P,22}^2 & = & \l (B_\mathrm{eff}+3\k s +\mu^\prime)\frac{v_u v_d}{s} -3\k A_\k s  -2 m_{S}'^2 -\k \mu^\prime s
-\xi_F\left(4\k + \frac{\mu^\prime}{s}\right) -\frac{\xi_S}{s}\; , \nn\\
{\cal M}_{P,12}^2 & = &\l (A_\l - 2\k s - \mu^\prime)\, v\; ,
\eea
where the combinations are defined as 
\beq
B_\mathrm{eff} = A_\lambda+ \kappa s\,,
\quad \widehat{m}_3^2 = m_3^2 + \lambda(\mu^\prime s + \xi_F)\,.
\eeq
Note that in the absence of mixing ${\cal M}_{P,11}$ gives the mass of the usual CP-odd field $A$ in the MSSM. Recall that for the purposes of this
discussion, we have assumed $m_3^2=\xi_F=0$, and now, for further simplicity, we will also assume that the combination $2 m_{S}'^2 +\frac {\xi_S}{s}=0$.
With these assumptions, the only new parameters present in this matrix but irrelevant for Higgsino-singlino mixing are the $A$-terms, $A_\lambda$ and $A_\kappa$. With these assumptions, we can set $\theta_a=0.1$ and fix the smaller eigenvalue of this matrix to be $m_a^2 \simeq (120 \rm GeV)^2$; we then obtain numerical values for the soft parameters $A_{\lambda,\kappa}$ and the mass of the heavier CP-odd state $A$. Since $\sin \theta_a$ is small, this state is similar in nature to the MSSM pseudoscalar boson. It is therefore constrained by searches for resonant di-tau production\cite{cmstautau,atlastautau}, and also potentially by flavor constraints\cite{bsmumu} such as those arising from measurements of the branching fraction $BR(B_s\to \mu^+\mu^-)$, for the large values of $\tan \beta$ ($\sim 40$) considered here. We therefore avoid these constraints by requiring $M_A > 1$ TeV. This approach works 
quite well and multiple solutions are again easily obtainable. 

Lastly, we turn to the CP-even Higgs mass-squared matrix, where we provide the tree-level elements here for completeness\cite{ellrev}: 
\bea
{\cal M}_{S,11}^2 & = & g^2 v_d^2 + (\mu_\mathrm{eff}\, B_\mathrm{eff} +
\widehat{m}_3^2)\,\tan\beta\;, \nn\\
{\cal M}_{S,22}^2 & = & g^2 v_u^2 + (\mu_\mathrm{eff}\, B_\mathrm{eff} +
\widehat{m}_3^2)/\tan\beta\;, \nn\\
{\cal M}_{S,33}^2 & = & \l (A_\l + \mu^\prime) \frac{v_u v_d}{s}
+ \k s (A_\k + 4\k s+ 3 \mu^\prime) - (\xi_S + \xi_F \mu^\prime)/s\;, \nn\\
{\cal M}_{S,12}^2 & = & (2\l^2 - g^2) v_u v_d -
\mu_\mathrm{eff}\, B_\mathrm{eff} - \widehat{m}_3^2 \;, \nn\\
{\cal M}_{S,13}^2 & = & \l (2 \mu_\mathrm{eff}\, v_d -
(B_\mathrm{eff} + \k s + \mu^\prime)v_u)\;, \nn\\
{\cal M}_{S,23}^2 & = & \l (2 \mu_\mathrm{eff}\, v_u -
(B_\mathrm{eff} + \k s + \mu^\prime)v_d)\;.
\eea
Here $g^2 \equiv (g_1^2+g_2^2)/2$ with $g_{1,2}$ being the usual SM gauge couplings. In order to make contact with 
experiment we must incorporate the important large stop/top and sbottom/bottom 1- and 2-loop radiative corrections to this matrix as given 
in \cite{ellrev}; this is particularly important for the $11, 22$ and $23$ elements. 
Given our assumptions outlined above, $m_{S}'^2$ is the only new free parameter entering this matrix that we need to scan. Here we employ the range $0 \leq m_{S}'^2 \leq s^2$, which is specifically chosen to avoid 
any tachyonic solutions for the eigenvalues of this matrix, for our scan. We require the lightest eigenvalue, which we identify 
as the Higgs boson observed at the LHC, to have a mass $m_h=125.5\pm 3$ GeV. We further demand that this state must have very little mixing with the singlet Higgs, \ie, $|S_{13}|^2 << 1$ in the notation of 
Ref.~\cite{ellrev}. This requirement ensures that the observed Higgs state will have properties that are very close to that of the MSSM and SM Higgs 
and, in particular, will have a small branching fraction for the decay $h\to \chi\chi$ (generally $\sim 1\%$), which can  
only proceed through the various neutralino and Higgs mixings. This also reduces the observed Higgs boson's contribution to the Spin-Independent Direct Detection cross section for the almost pure singlino LSP.  

Performing a scan over the parameters ${\lambda,\kappa, t_\beta, s, m_{S}'^2}$ yields many sets of parameter values (\ie, `models') that produce 
completely successful phenomenology. Certainly allowing the other remaining parameters that we have ignored in our discussion to be non-zero or to float 
freely will also yield viable scenarios.  In addition, increasing the LSP mass to 70 GeV will clearly still allow for solutions.
Thus, at least at (mostly) tree level, the general NMSSM can provide a phenomenologically viable 
description of the FGCE!

\subsection{Full Scan of the General NMSSM}

To go further in our exploration of the NMSSM we perform a more thorough and complete analysis incorporating, \eg,  the higher-order, 
loop-generated radiative corrections to the various gaugino, CP-even and CP-odd Higgs mass matrices as well as the associated couplings in the NMSSM. 
In performing this analysis we make use of a modified version of {\tt NMSSMTools4.3.0}\cite{Tools,Ellwanger:2004xm}{\footnote {We have made a 
number of modifications 
to the default vanilla version of {\tt NMSSMTools4.3.0} for the analysis presented here; these are described in detail below.}} and perform a scan over the 
general NMSSM parameter space. We subject each of the randomly chosen parameter space points to the various experimental 
and theoretical constraints mentioned above, along with additional constraints which will be detailed in the next section, in order to find 
phenomenologically successful models that describe both the observed relic density and the FGCE.

As was noted in the previous discussion, the general NMSSM contains $10$ new parameters beyond those already present in the MSSM which we now include 
as part of our scan. An outline of our scan procedure is as follows.
For simplicity, we choose fixed values for the usual gaugino masses ($M_{1,2,3}$), the slepton soft masses, and the leptonic $A$-terms, and scan over the ratio of the two Higgs vevs, $\tan \beta$. Additionally, we set the NMSSM parameters $m_{S}'^2=m_3^2=0$ since they play no important role in this analysis. Motivated by the FGCE measurements and our previous discussion, we are specifically interested in particular ranges for the physical masses of the LSP, the lightest CP-odd scalar and the lightest CP-even scalar (which we equate with the $\sim 125$ GeV Higgs as discussed above). We also want to constrain the mass of the heavier CP-odd field from below to satisfy constraints from LHC searches and flavor physics. 
We therefore eschew scanning over the fundamental Lagrangian parameters $\mu^\prime$ and $\xi_{F,S}$ in favor of a scan over the interesting ranges of 
the physical masses $m_a$, $m_A$, $m_h$ and $m_{LSP}$.  The (common) squark masses and the corresponding squark $A$-terms are also 
set by choosing values for these physical masses, since suitably large radiative corrections are necessary in the NMSSM (as in the MSSM) in order to obtain a Higgs mass 
in the $\sim 125$ GeV range. After performing this substitution, the remaining scan parameters are just $\lambda, \kappa, \mu_{eff}$ 
and the soft parameters $ A_{\lambda,\kappa}$. The approach that we have just outlined mimics the procedure that we followed in our tree-level discussion as closely as possible.  

We now discuss our scan parameters and their ranges in a bit more detail. Some of the parameters are more relevant to our aim of demonstrating the existence of points that  
can explain the FGCE than others.  We therefore divide the set of parameters into ``fixed parameters,'' which are 
not expected to have a significant effect on the 
relevant physics and hence are set to reasonable values, and ``scanned parameters,'' which are 
scanned using a uniform distribution (\ie, employing flat priors) within the ranges specified below.  In {\tt NMSSMTools}, these parameters are defined at the scale 
$Q^2 = (2 m^2_{\tilde{Q}_1} + m^2_{\tilde{u}_1} + m^2_{\tilde{d}_1})/4$ and are then run to $Q^2 = m_{\tilde{Q}_3}m_{\tilde{u}_3}$ (with the exception of $\tan{\beta}$, 
which is input at $Q = M_Z$).  Since we choose a common squark mass parameter, $m_{\tilde Q}=m_{(\tilde{u},\tilde{d})_{1,2,3}}$, for our analysis, these scales are both equal to 
$m_{\tilde Q}$. These scales are slightly different from the geometric mean of the stop masses, which is often employed as the SUSY scale, but we 
do not expect this difference to have any effect on our results.
\smallskip

\noindent\textbf{Fixed Parameters:}  
As the gauginos do not play an important role in our scenario, provided they are sufficiently heavy to evade any experimental constraints, we set 
\begin{equation}
\label{eq:gaugino-masses}
2 M_1 = M_2 = M_3 /3 = 1\, \rm{TeV}\,.
\end{equation}
Similarly we fix the slepton masses to be
\begin{equation}
\label{eq:slepton-masses}
m_{\tilde{L}_1} = m_{\tilde{L}_2} = m_{\tilde{L}_3}
= m_{\tilde{e}_1} = m_{\tilde{e}_2} = m_{\tilde{e}_3} = 1\, \rm{TeV}\,,
\end{equation}
and the third generation lepton trilinear coupling is set to
\begin{equation}
\label{eq:Atau}
A_\tau = 1.5\, \rm{TeV}.
\end{equation}
We do not expect our results to be particularly sensitive to any of these choices. 
Additionally, in the general NMSSM, one could assign non-zero values to the Lagrangian parameters $m_3^2$ and 
$m_{S^\prime}^2$.  However, as noted above, we have fixed 
these values to zero at the input scale. 
\smallskip

\noindent\textbf{Scanned Parameters:}
The remaining Lagrangian parameters are chosen via a scanning procedure.  These parameters, together with the 
ranges from which they were chosen using uniform 
distributions (flat priors), are listed in Table~\ref{tab:lagrangian-parameters}. Note that we perform a much 
broader scan of the parameters than for our
tree-level study described above, since we are now interested in characterizing the phenomenologically interesting region in addition to demonstrating its existence.


\begin{table}[t]
\centering
\begin{tabular}{| c | c | c  | c |}
\hline
\hline
Parameter & Value  & Lower Bound & Upper Bound \\ 
\hline \hline
$M_1$                      & $500$ GeV     &  ---  &   --- \\
$M_2$                      & $1$ TeV      &  ---  &   --- \\
$M_3$                      & $3$ TeV      &  ---  &   ---  \\
$m_{\tilde L(\tilde{e})_{1,2,3}}$ & $1$ TeV       &  ---  &   --- \\ 
$m_3^2$                  & $0$      &  ---  &   ---  \\
$m_{S^\prime}^2$       & $0$     &  ---  &   --- \\
$A_{\tau}$                 & $1.5$ TeV      &  ---  & --- \\
$\tan \beta$           & Scanned  &  $1$        & $60$     \\
$\lambda$              & Scanned  &  $0$          & $0.7$    \\        
$\kappa$                & Scanned  &  $-0.7$     & $0.7$     \\   
$A_\lambda$             & Scanned  &  $-30$ TeV & $30$ TeV \\   
$A_\kappa$              & Scanned  &  $-30$ TeV & $30$ TeV \\      
$\mu_\mathrm{eff}$  & Scanned &  $-5$ TeV   &  $5$ TeV  \\ 
$m_{\tilde Q}$          & Replaced & ---   & ---  \\ 
$A_{t,b}$                  & Replaced & ---   & ---  \\
$\xi_F$                   & Replaced & ---   & ---  \\
$\xi_S$                   & Replaced & ---   & ---  \\
$\mu^\prime$          & Replaced & ---   & ---  \\
\hline
\hline
\end{tabular}
\caption{Summary of how the parameters of the general 
NMSSM were chosen in our scan.  The ``Value'' column gives the parameter value for fixed parameters. A ``Scanned'' entry in this column indicates that the parameter value was chosen randomly from a uniform distribution within the range specified by the subsequent two columns. A ``Replaced'' entry indicates that the parameter value is solved for numerically to obtain desired values of the physical parameters listed in 
Table~\ref{tab:physical-parameters}.}
\label{tab:lagrangian-parameters}
\end{table}


As noted in Table~\ref{tab:lagrangian-parameters}, we have substituted several of the Lagrangian parameters with the corresponding physical parameters $m_h$, $m_{a}$, 
$m_{A}$, and $m_{\chi_1^0}$.  This increases the efficiency of our scan since we are only interested in a
restricted range of values for $m_h$, $m_{a}$, and 
$m_{\chi_1^0}$.  In addition, $m_{A}$ is generically strongly constrained by experiment and is not allowed to be small.
This efficiency boost is particularly necessary since we only obtain acceptable models at somewhat finely tuned regions of the parameter space, with a significant cancellation between $\mu'$ and $2 \kappa s$ ($= \frac{2 \kappa \mu_{\rm{eff}}}{\lambda}$ using our scan parameters). As we are interested in parameter space points for which both of these terms are $\mathcal{O}$(TeV), yet $30$ GeV $< |m_{\chi_1^0}| < 40$ GeV, it is obvious that scanning 
over all the parameters $\kappa, \lambda, \mu_{\rm{eff}}$, and $\mu^\prime$ will be inefficient. 

In order to proceed, we first set all squark mass parameters to a common value:
\begin{equation}
  \label{eq:squarks}
  m_{\tilde Q}\equiv m_{\tilde{Q}_1} = m_{\tilde{Q}_2} = m_{\tilde{Q}_3}
  = m_{\tilde{u}_1} = m_{\tilde{u}_2} = m_{\tilde{u}_3}  
  = m_{\tilde{d}_1} = m_{\tilde{d}_2} = m_{\tilde{d}_3},
\end{equation}
as noted above, and the third generation squark trilinear couplings to be proportional to the squark mass parameter
\begin{equation}
  \label{eq:At-Ab}
  A_t = A_b = \sqrt{6} m_{\tilde Q}\,.
\end{equation}
This expression was used to facilitate large stop mixing, providing large positive corrections to the lightest Higgs mass. 
We employed the {\tt NMSSMTools} default functionality to replace $\xi_F$ and $\xi_S$ with the variables {\tt MP} and {\tt MA}, which represent the first and second diagonal elements, 
respectively, of the effective $2 \times 2$ CP-odd Higgs mass matrix (\ie, with the Goldstone boson removed) at tree level.

To begin our scan, we chose target values of $m_h$, $m_{a}$, $m_{A}$, and $m_{\chi_1^0}$, along with values for the remaining scanned parameters. To find the Lagrangian parameters ($m_{\tilde Q}$, {\tt MP}, {\tt MA}, $\mu^\prime$) corresponding to these target values, we chose a seed point in the ($m_{\tilde Q}$, {\tt MP}, {\tt MA}, $\mu^\prime$) parameter space that would be unlikely to lead to a tachyonic particle in the spectrum. We then used a simple variant of Newton's method to numerically search for a point giving the correct physical masses. In this search method, the `next step' in parameter space was scaled by a small constant to reduce the chances of obtaining a tachyonic spectrum. If the numerical search for appropriate physical masses failed, a new ``safe guess'' for the initial parameter values was made and the procedure restarted. 
However, only a finite number of such ``restarts'' were allowed.  If too many searches ended unsuccessfully as a result of, \eg,  
any of the particles being tachyonic, or if the search failed to obtain a point with the 
required values for the physical masses after a large number of iterations, the scan point was rejected.

%

\begin{table}[t]
\centering
\begin{tabular}{| c | c | c  | c |}
\hline
\hline
Parameter & Value  & Lower Bound & Upper Bound \\ 
\hline \hline
$m_h$                      & Scanned      &  $122$ GeV  &   $128$ GeV \\
$m_{a}$                  & Scanned      &  $80$   GeV  &   $800$ GeV  \\
$m_{A}$                  & Scanned      &  $500$ GeV  &   $5$ TeV  \\
$|m_{\chi_1^0}|$          & Scanned      &  $30$   GeV  &   $40$ GeV \\ 
\hline
\hline
\end{tabular}
\caption{Ranges employed in our scan for the physical values of the listed parameters.
Here $m_h$ is the mass of the lightest CP-even Higgs boson, $m_{a,(A)}$ is the lighter (heavier) 
CP-odd Higgs boson, and $m_{\chi^0_1}$ is the signed mass of the lightest neutralino, which is also the LSP.  
We scan over both signs of the LSP mass.  
Again note that these enlarged ranges allow for a full exploration of the 
phenomenologically interesting parameter regimes.}
\label{tab:physical-parameters}
\end{table}


In performing this general parameter scan a number of issues arose that required 
modifications of the default {\tt NMSSMTools4.3.0} package and that partially 
influenced the constraints we have applied in exploring the NMSSM parameter space. For example, as 
noted above, the calculation of Higgs masses in the {\tt NMSSMTools} 
package includes radiative corrections to both the scalar and pseudoscalar neutral Higgs mass matrices, as 
well as to the charged Higgs mass.  We find that for much of 
the parameter space relevant to our study, the radiative correction to the $(1,2)$ element of the neutral scalar 
Higgs mass matrix arising from chargino loops lead to 
spuriously large corrections, resulting in a tachyonic Higgs sector.  This is most likely due to a failure to properly resum the large logarithms. 
In order to remove this unphysical effect in a 
consistent manner, we neglected the radiative corrections from chargino loops to the scalar and pseudoscalar 
mass matrices, as well as to the charged Higgs mass, if 
in \textit{any} of these matrices the absolute value of the modification of any entry arising from the application of this 
specific radiative correction was greater than the 
initial absolute value of that entry.  None of the other radiative corrections included in {\tt NMSSMTools} were 
altered in any way as in these cases the corrections were under control.

Furthermore, {\tt NMSSMTools} employs {\tt micrOMEGAs}\cite{Belanger:2005kh, Belanger:2006is, Belanger:2008sj, Belanger:2010gh, Belanger:2013oya} 
to calculate the thermal relic density and other dark matter properties.  As {\tt micrOMEGAs} is based on {\tt CalcHEP}~\cite{Belyaev:2012qa}, 
a description of all the vertices in the theory under consideration 
can be found in the appropriate model file. In this file we found that the Higgs-neutralino-neutralino couplings 
are written in terms of neutralino masses.  This is acceptable in the $Z_3$-invariant NMSSM, but in the general NMSSM we found, \eg, an additional 
contribution to the singlino mass from the $\mu^\prime$ term, as shown
above.  This term \textit{does not} contribute to the Higgs coupling with the singlino.  This is extremely important 
for our purposes as we are considering parameter space regions 
with a relatively light singlino LSP, but with an unsuppressed singlet-singlino-singlino coupling.  Therefore we 
needed to replace the expressions for the $h \chi_1^0 \chi_1^0$ and $a \chi_1^0 \chi_1^0 $ vertices in the relevant model file with those found in 
Ref.~\cite{ellrev}.  In implementing the expressions from \cite{ellrev}, 
we were careful to choose signs such that the new vertex  
expressions agree with the previous ones in the limit of the $Z_3$-invariant NMSSM.  To repeat, in order to correct for this error, we have replaced 
the LSP couplings to the two lightest CP-even and CP-odd scalars in {\tt micrOMEGAs} with the appropriate expressions 
from Ref~\cite{ellrev}. We note that these couplings determine the dominant contributions to both the indirect and direct detection cross sections in the scenarios considered here.

\subsubsection{Theoretical Constraints Applied to the Parameter Space}

In performing this analysis, we generated a set of $6\cdot 10^8$ points in the NMSSM parameter space 
according to the procedure discussed above and passed them through 
various experimental and theoretical constraints. Some of the details are worth a short description.  

To ensure theoretical consistency of our models, we remove those that are flagged by {\tt NMSSMTools} as 
having any of the following problems: ($i$) Negative mass-squared values 
for the heavy scalar Higgses, pseudoscalar Higgses, or sfermions, ($ii$) zero values for $\tan \beta$, $\lambda$, or $\mu$, ($iii$) numerical problems in the spectrum calculation and ($iv$) the presence of an unphysical minimum in the scalar potential. 
In addition to these constraints, we also check for cases where the radiative corrections are potentially non-perturbative. 
One example of this is the chargino corrections to the Higgs couplings, described above. Another case in which these couplings can become extremely large is in the 
couplings of the Higgses to bottom quarks. Here, it is possible to generate extremely large radiative corrections when $\mu_{eff}$ is large and negative and $\tan \beta$ is also large. 
This type of correction is well known, and can in fact be an O(1) 
effect, as described in \cite{Cahill-Rowley:2014wba}. The size of this correction for the scalar $\phi_i$ can 
be estimated by taking the ratio
\begin{equation}
r_i=\frac{\Gamma(\phi_i \to b\bar{b})}{\Gamma(\phi_i \to \tau^+ \tau^-)}\,.
\end{equation}

Although the tree-level prediction for these $r_i$ is approximately 10, a sizable set of models have $r_i$ 
above 100 for one or more Higgses, suggesting that the corrections 
may need to be resummed. Since this resummation is beyond the scope of this work, we simply discarded 
models with potentially unphysical corrections. We estimated the range of 
$r_i$ that might result from valid corrections by examining our set of pMSSM models whose Higgs properties 
are studied in \cite{Cahill-Rowley:2014wba}, and for which the $hb\bar{b}$ 
coupling corrections can be large but are under control. We define the corresponding minimal and maximal values 
according to the extremal values appearing in the pMSSM model set as follows:
\begin{equation}
r_{max}=\frac{\mathrm{Max}(\Gamma(h \to b\bar{b}))}{\mathrm{Min}(\Gamma(h \to \tau^+ \tau^-))}\,,\quad\quad\quad\quad
r_{min}=\frac{\mathrm{Min}(\Gamma(h \to b\bar{b}))}{\mathrm{Max}(\Gamma(h \to \tau^+ \tau^-))}\,.
\end{equation}

With these definitions, we find $r_{max} = 60.0$ and $r_{min} = 2.42$, which we then use as upper and lower 
bounds on $r_i$ in the NMSSM.  We note that only very large corrections to $r_i$  
modify our results significantly, and that our results are therefore insensitive to the precise values of 
$r_{max}$ and $r_{min}$ that we choose.

\subsubsection{Experimental Constraints Applied to the Parameter Space}

\begin{table}[t]
\hskip-1.45cm\begin{tabular}{| c | c | c  | c |}
\hline
\hline
Observable  & Experiment & SM Prediction & Estimated Uncertainty \\ 
\hline \hline
$\Delta M_{B_d}$[${\rm ps}^{-1}$]           & {$ 0.507 \pm 0.005$} {\cite{Amhis:2012bh}}     &  0.545 $\pm$ 16.8\% {\cite{pdg} \cite{hfag}} &   Parameter dependent  \\
$\Delta M_{B_s}$[${\rm ps}^{-1}$]           & {{\small $17.719\pm 0.043$}}  {\cite{Amhis:2012bh}}      &  17.70 $\pm$ 6.3\%  {\cite{pdg} \cite{hfag}} &   Parameter dependent  \\
BR($B_s \to \mu^+ \mu^-$) $\times 10^{9}$                   & {$2.9 \pm 0.7$} {\cite{bsmumu}}     &  3.74 $\pm$ 4.1 \%  {\cite{pdg} \cite{hfag}} &   Parameter dependent \\
BR$(B \to X_s \, \gamma)$ $\times 10^{4}$   & {$3.43 \pm 6.7$}  {\cite{Amhis:2012bh}}    &  3.17 $\pm$ 7.3\%  {\cite{pdg} \cite{hfag}}&   Parameter dependent  \\ 
BR($B\to \tau \nu_\tau$) $\times 10^{4}$                & {$1.14 \pm 0.22$} {\cite{Amhis:2012bh}}      &  0.779 $\pm$ 8.6\%  {\cite{pdg} \cite{hfag}} &   Parameter dependent  \\

$\Omega h^2$  & $0.119 \pm 0.006$ \cite{Ade:2013zuv}      &  --- &   $\pm 10\%$ \\
$\sigma_{SI,Xenon}$     & $\sigma - m_\chi$ plane \cite{Akerib:2013tjd}     &  --- &   $\pm 400\%$ \\

$m_h$ (GeV)    & $125.03^{+0.39}_{-0.42}$ \cite{CMS:2014ega}   &  --- &   $\pm 3$ \\
$BR(h\to \chi \chi)$+$BR(h\to b\bar{b})$  & 0.468-0.686{\footnote{Not technically a measured quantity, but derived from the observed signal strength (see text for details)}}   &  0.575 &   --- \\

$\Gamma(Z \to \chi \chi)$      & \textless $ 1.7 \cdot 10^{-3}$ \cite{Tools}     &  0 &  --- \\
$\phi \to \tau \tau$    & Process dependent \cite{cmstautau,atlastautau}      &  0 &   --- \\
$\sigma_{NP},LEP$     & Process dependent\cite{pdg}      &  0 &  --- \\
$\sigma_{NP},LHC$     & Process dependent\cite{pdg}      &  0 &  --- \\	

\hline
\hline
\end{tabular}
\caption{List of the experimental constraints applied to the NMSSM models. The Experiment column 
gives the experimental central values and errors used in implementing the 
constraints. The SM Prediction column denotes the SM predictions for the observables (if applicable). The Estimated Uncertainty column lists the estimated theory uncertainty for the value of the observable as calculated by {\tt NMSSMTools}.}
\label{tab:expcon}
\end{table}

To create a selection of viable NMSSM models, we apply the set of experimental bounds listed in Table~\ref{tab:expcon}. 
These can be divided into constraints from flavor physics, 
Higgs measurements, dark matter observables, and collider searches. The first 5 observables in the Table are flavor 
observables that are computed by {\tt NMSSMTools}. This calculation  
includes an estimate of the theoretical uncertainty, including both parametric errors and uncertainties from 
missing terms in the diagrammatic expansion. For a given model, the combined 
uncertainty is the linear sum of the (model-specific) theory uncertainty and the experimental error. We designate a model as being viable if it agrees with the experimental value within 
this combined uncertainty. Since the experimental ranges in {\tt NMSSMTools} are supplanted by more recent results for the flavor observables that we consider, we have modified {\tt NMSSMTools} to use the updated experimental results listed in Table~\ref{tab:expcon}. We also note that the theoretical precision of flavor observable calculations within {\tt NMSSMTools} does not match that of the most precise SM calculations, with the result that the SM value predicted by {\tt NMSSMTools} can differ from the SM values listed in Table~\ref{tab:expcon}. For the purposes of determining which models are excluded by flavor measurements, this fact is included in the estimated theoretical uncertainty. In our discussion, however, we will normalize 
the {\tt NMSSMTools} prediction for each observable by the corresponding SM value \textit{as predicted by {\tt NMSSMTools}} in order to clearly show the effects which result from BSM physics, and not from e.g. different values for the CKM parameters. Although {\tt NMSSMTools} computes the NMSSM contributions to the 
muon magnetic moment, we do not use this observable as a constraint given the unclear 
status of the relevant theoretical uncertainties and the experimental agreement with the SM. We note, however, 
that all of the models in our final selection are within the range ($-17.7 \cdot 10^{-10} - 43.8 \cdot 10^{-10}$) for $\Delta(g-2)_\mu$, which is
the region we allowed in our study of the pMSSM\cite{uspmssm}.

As noted above, we assume that the LSP is a thermal relic and accounts for all of the observed dark matter abundance. 
As a result, the relic density measurement from Planck and 
the null results from direct detection experiments place important constraints on our models. We employ the default 
treatment of the relic density within {\tt NMSSMTools}, requiring 
it to be within 10\% of the Planck observation to account for theoretical uncertainties in the relic density calculation 
(the experimental errors are negligible by comparison). 
For direct detection, we use the LUX limit on the spin-independent cross section, but rescale it by a factor of 4 to 
account for matrix element uncertainties in the nucleon, \eg, the strange content 
of the proton.  Including such uncertainties has minimal impact on our results. Spin-dependent experiments 
are not yet sensitive to models that remain allowed by the other constraints, so we make no explicit requirement on 
the spin-dependent cross section.

Constraints from observations of the SM-like Higgs boson also set important limits on our model parameters. First, as 
noted above, we require that the lightest Higgs mass  
be in the range 122-128 GeV, assuming a 3 GeV theoretical uncertainty in the Higgs mass calculation\cite{Higgsmass} with the  
experimental value of $\sim$ 125 GeV.\footnote{We note the present embarrassing state wherein the errors associated with the
theoretical computation of the Higgs mass are much larger, by approximately an order of magnitude, than the experimental
uncertainty.}  Secondly, we note that 
the couplings of the 125 GeV Higgs to SM final states (other than $b\bar{b}$) are very close to their SM values in the 
models we consider. Since the $b\bar{b}$ channel 
is presently poorly constrained by direct measurements, the most important effect of increasing (decreasing) the 
$hb\bar{b}$ coupling in our model set is to simultaneously reduce (enhance) 
the Higgs branching fractions to the other SM final states, thus altering the total observed signal strength. This same effect also 
provides the dominant constraint on the Higgs invisible 
branching fraction, given the SM-like nature of the light Higgs couplings. Reproducing the CMS measurement of 
$\mu=1 \pm 0.26$ (\ie, $2\sigma$ confidence), where $\mu$ denotes the relative strength of the measured Higgs couplings to their SM
expectations averaged over all final states, requires that the 
sum of the branching fractions for $b\bar{b}$ and $\chi \chi$ final states lies between 0.468 and 0.686.  Here, we neglect 
the direct observation of the $b\bar{b}$ final state, since 
the presently large errors on this measurement make this a reasonable approximation.

Finally, we apply constraints arising from the non-observation of SUSY particles at LEP and the LHC. These restrictions 
include an upper limit on the invisible width of the $Z$ boson from LEP, 
bounds on superpartner masses from both LEP and LHC searches, and direct searches for scalar resonances decaying to ditau 
final states at the LHC. Given our scan ranges for the NMSSM parameters, only 
charginos, stops and sbottoms can {\it a priori} be in tension with either the LEP or LHC data. Since the masses of these 
sparticles are generally unimportant for the 
dark matter scenario we consider here, we simply avoid these collider searches by requiring charginos to be heavier than 100 
GeV and stops/sbottoms to be heavier than 700 GeV. 
Finally, we make use of the {\tt NMSSMTools} constraints on the ditau signal strength in the gluon fusion and $b\bar{b}$-associated 
production channels, which are derived by comparing the 
calculated signal strengths to the limits given in \cite{cmstautau}.

After applying these experimental constraints, we make the final requirement that our models provide a good explanation of 
the galactic center excess. As noted above, we require 
a LSP mass in the 30-40 GeV range (set by our choice of scan range) and then further require that the present-day annihilation 
rate exceeds $0.7 \cdot 10^{-26}$~cm$^3$s$^{-1}$ in order 
to obtain the required FGCE described by Ref.~\cite{recentgce}. All of our models will necessarily predict a present-day annihilation 
cross section which is slightly suppressed compared 
to that for the early universe as described earlier, and are therefore consistent with upper limits on the annihilation flux. 

After all of these constraints are applied, a set of $\sim 52.8$k NMSSM model points remain viable, whose properties we now discuss in some detail.

\section{NMSSM Numerical Results and Discussion}
    
\begin{figure}[htbp]
\centerline{\includegraphics[width=3.5in]{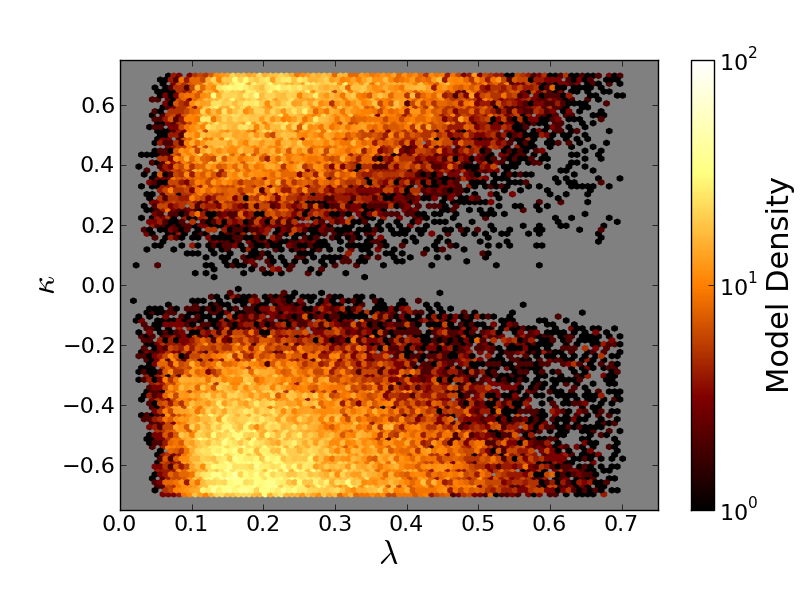}
\hspace{0.30cm}
\includegraphics[width=3.5in]{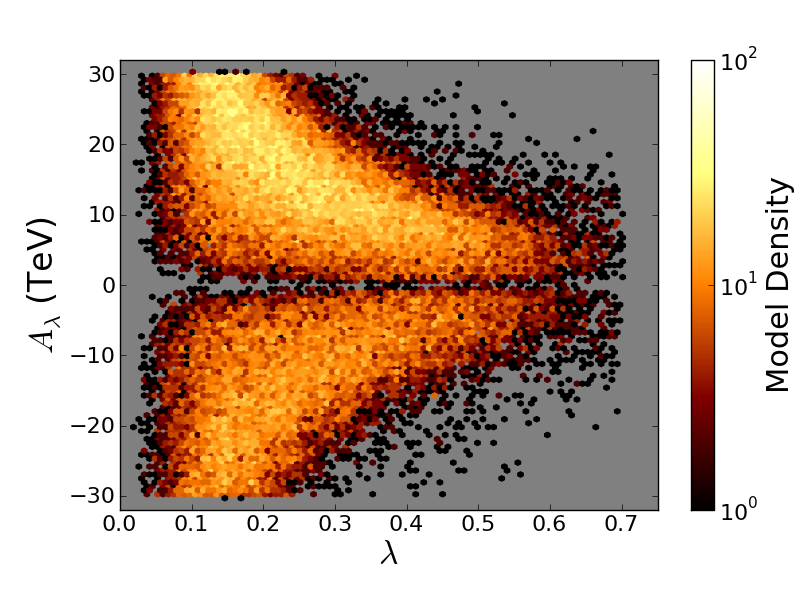}}
\vspace*{0.50cm}
\centerline{\includegraphics[width=3.5in]{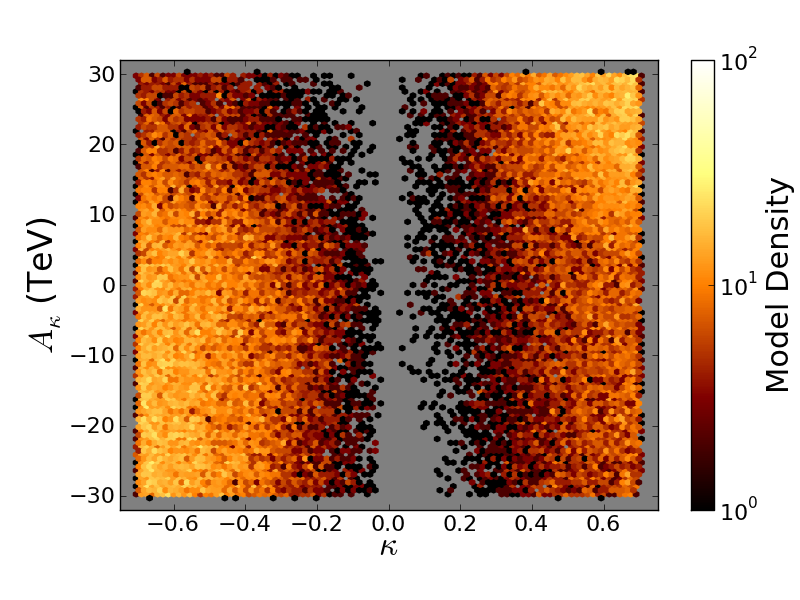}
\hspace{0.30cm}
\includegraphics[width=3.5in]{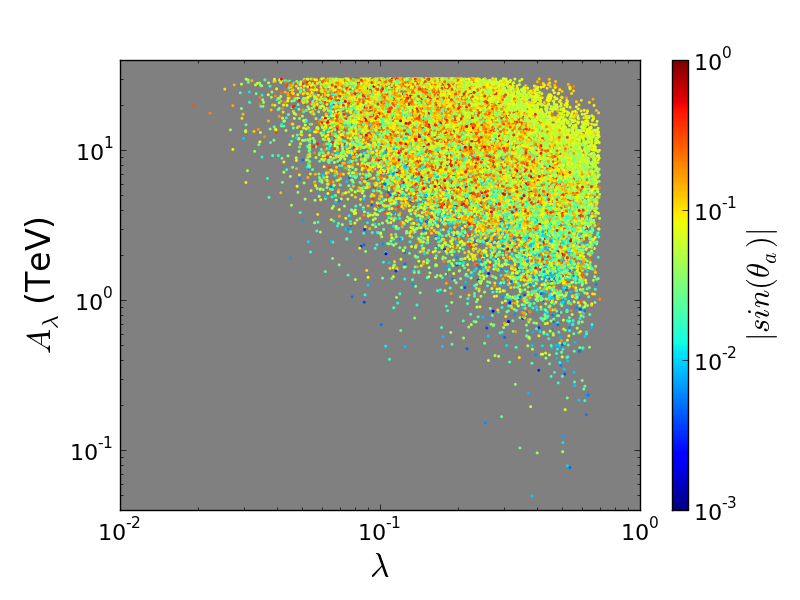}}
\vspace*{-0.10cm}
\caption{Two-dimensional projections of the NMSSM parameter space, depicting the 
allowed regions that describe the FGCE, color-coded by the density of allowed models in each
bin.  Top-left panel: $\kappa-\lambda$ plane. Top right: $A_\lambda - \lambda$ plane.
Bottom left: $A_\kappa-\kappa$ plane.  Bottom right: $|A_\lambda|-\lambda$ plane, color-coded by
the absolute value of the pseudoscalar mixing angle $\sin\theta_a$.}
\label{fig5}
\end{figure}

\begin{figure}[htbp]
\centerline{\includegraphics[width=3.5in]{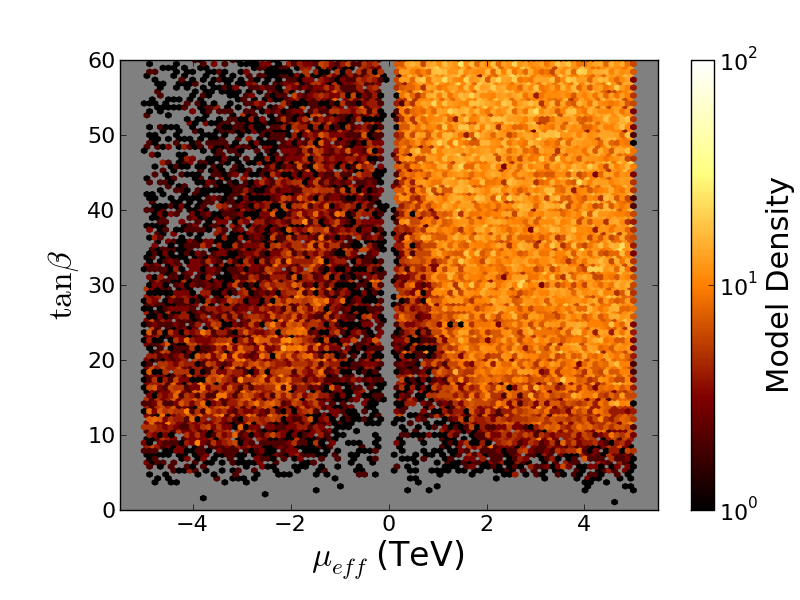}
\hspace{0.30cm}
\includegraphics[width=3.5in]{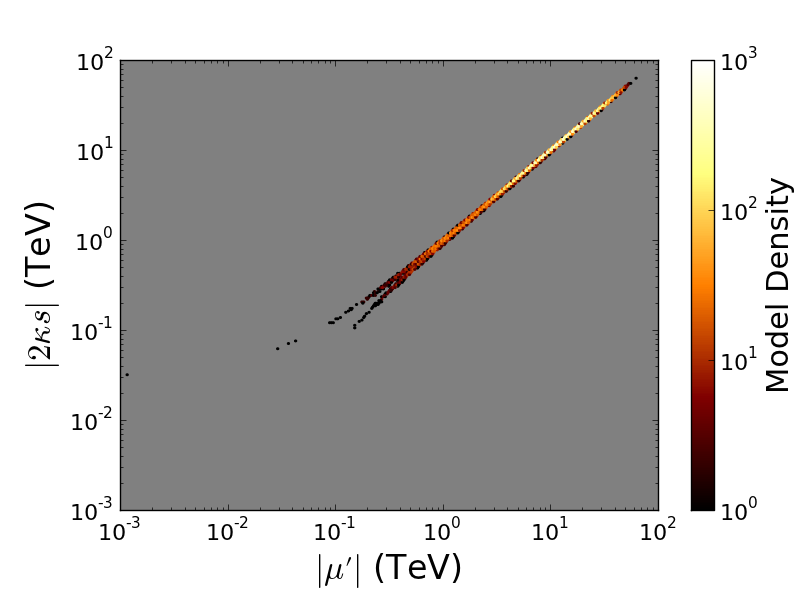}}
\vspace*{0.50cm}
\centerline{\includegraphics[width=3.5in]{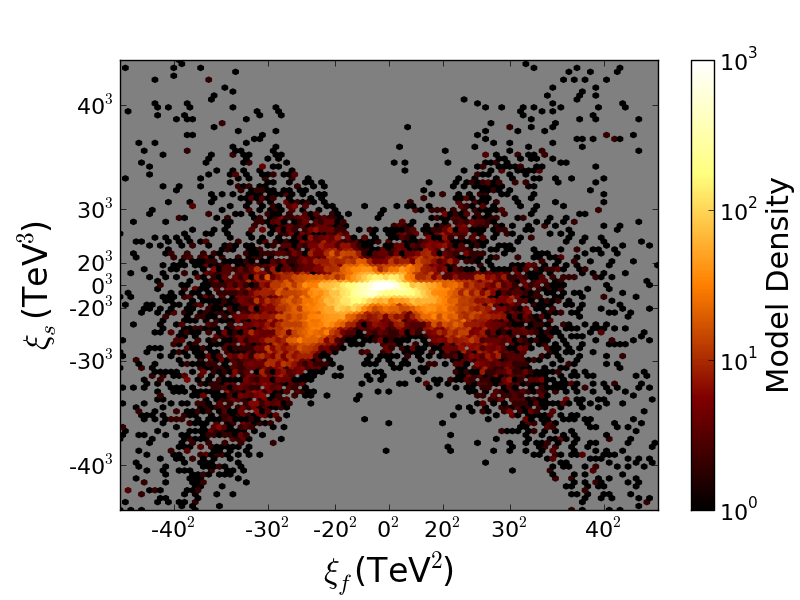}
\hspace{0.30cm}
\includegraphics[width=3.5in]{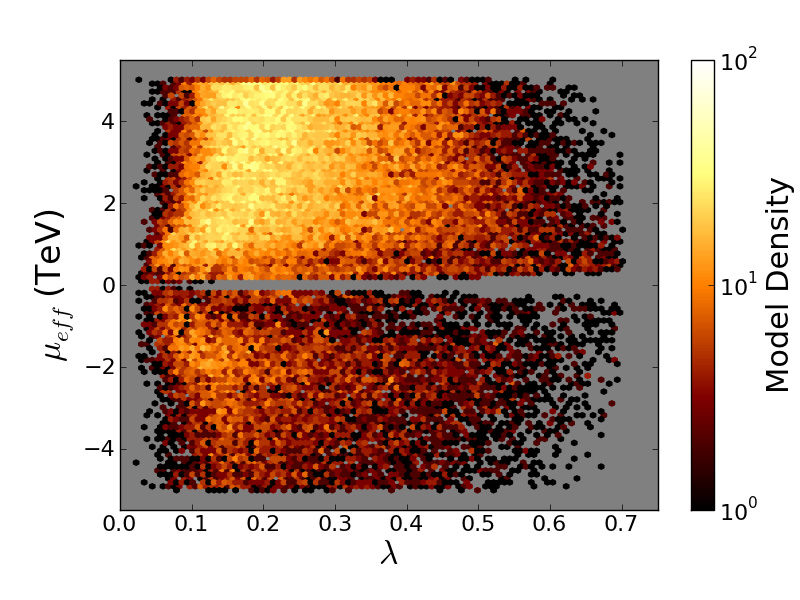}}
\vspace*{-0.10cm}
\caption{Two-dimensional projections of the NMSSM parameter space, depicting the 
allowed regions that describe the FGCE, color-coded by the density of allowed models in each
bin.  Top-left panel: $\tan\beta - \mu_{eff}$ plane. Top right: $|2\kappa s| - |\mu^\prime|$ plane.
Bottom left: $\xi_S - \xi_F$ plane.  Bottom right: $\mu_{eff}-\lambda$ plane.}
\label{fig6}
\end{figure}

We begin by examining the regions for the fundamental parameters of the superpotential and soft-breaking Lagrangian that reproduce
the FGCE and are consistent with all other constraints we have applied.
Recall that the values of both the 
superpotential and soft parameters were taken from 
initially flat distributions.  The first three panels of Fig.~\ref{fig5} and all four panels of Fig.~\ref{fig6} display the density of models in a region, as projected onto 
planes formed by pairs of these fundamental parameters. 
There are several observations to note: ($i$) The model density distribution for $\lambda$ reaches maximum near $\sim 0.15$ while $\kappa$ 
prefers to take on larger absolute values with its model density being asymmetric with respect to its sign.{\footnote {The $\kappa$ dependence of many physical quantities is quadratic and thus sign-independent.  However, the scalar and pseudoscalar mass matrices do depend on the sign of $\kappa$, and are likely to be the origin of this sign asymmetry.}}  ($ii$) For positive values of $\kappa$, we note that a reduction in model density appears when $\lambda$ 
is large and $\kappa$ is small; this does not occur for 
negative values of $\kappa$. ($iii$) The distribution of $A_\kappa$ is quite uniform over the chosen scan range while that 
of $A_\lambda$ peaks below $\sim 10$ TeV and is seen to 
be sign-asymmetric with positive values being preferred. Note that since $A_\lambda$ always appears together with $\lambda$ 
itself in the soft-breaking Lagrangian, their values are correlated so that the product 
$|\lambda A_\lambda|$ always lies below $\sim 15$ TeV.  ($iv$) The density of 
$\tan \beta$ values is relatively constant above $\sim 10$ but on rare occasions models are seen to extend down to as 
low as $\tan \beta \sim 1.5$. For large $\tan \beta$, the 
set of models with large negative values of $\mu_{eff}$ are seen to be relatively depleted. ($v$) Unsurprisingly the 
values of $2\kappa s$ and $\mu^\prime$ are very highly 
correlated as their sum at tree level produces the rather small LSP mass (which may take either sign). 
While $|\mu^\prime|$ tends to have values of several TeV or more (with 
better than $1\%$ tuning in the LSP mass as described above), there is a long tail stretching down to much lower 
values, even below 100 GeV. There is, in fact, a single model in our set 
with $|\mu^\prime| \sim 1$ GeV wherein the LSP picks up a substantial Higgsino content and the mass of the $a$ is 
not far from $2m_\chi$. This case exhibits behavior similar to the $Z_3$-symmetric scenario, but is exceedingly 
fine tuned, with the relic density being generated by annihilation near the $Z$ pole while the FGCE is obtained through 
LSP pair annihilation through the $a$ pole, as in the model of 
\cite{nmssmkath}.  ($vi$) We note that the fundamental parameters $\xi_{(F,S)}$ have dimensions of mass$^{(2,3)}$, 
respectively, so that it is most useful to present their values as a 
mass in TeV$^{(2,3)}$. Although a sprinkling of models have values of $\xi_{F,S}$ that extend to both very large positive and negative mass$^{(2,3)}$, 
both of these distributions peak in the 
negative region with values of order a few TeV$^{(2,3)}$ as might be expected.
($vii$) Lastly, $\mu_{eff}$ is seen to slightly prefer positive values of a 
few TeV with the mass range $\lesssim 100$ GeV  
being excluded by the lower bound on the Higgsino mass from LEP. For positive values of $\mu_{eff}$ we see that slightly larger 
values of $\lambda$ are preferred.

\begin{figure}[htbp]
\centerline{\includegraphics[width=3.5in]{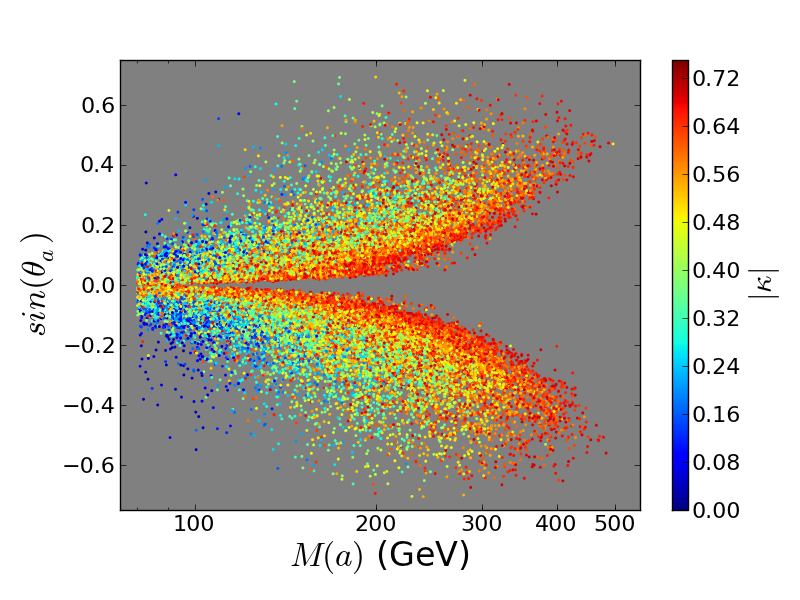}
\hspace{0.30cm}
\includegraphics[width=3.5in]{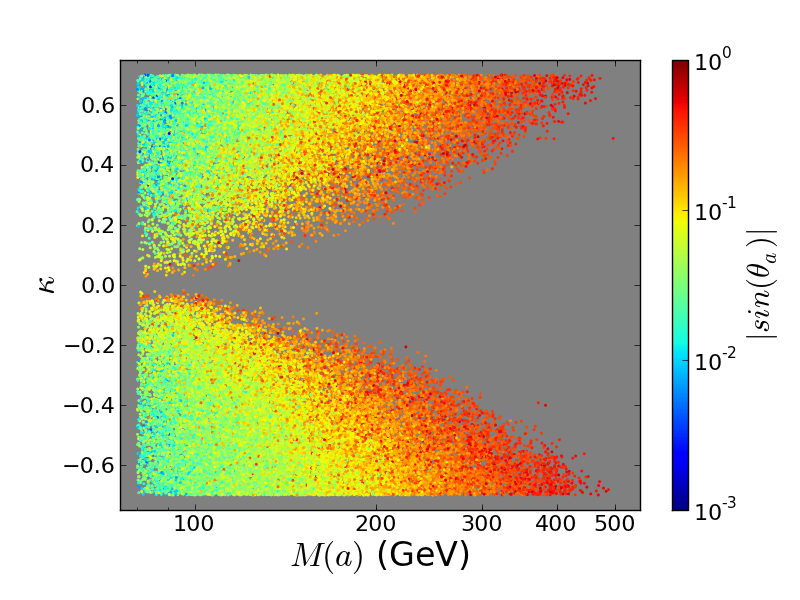}}
\vspace*{0.50cm}
\centerline{\includegraphics[width=3.5in]{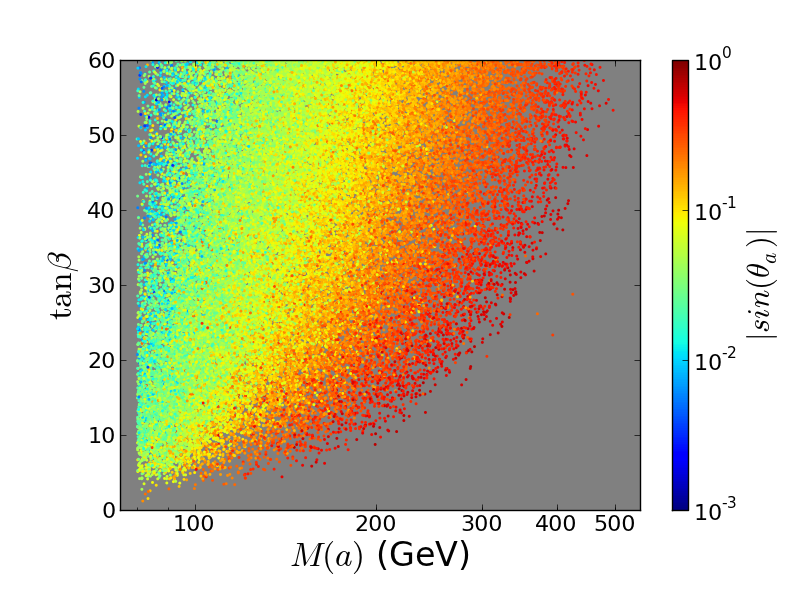}
\hspace{0.30cm}
\includegraphics[width=3.5in]{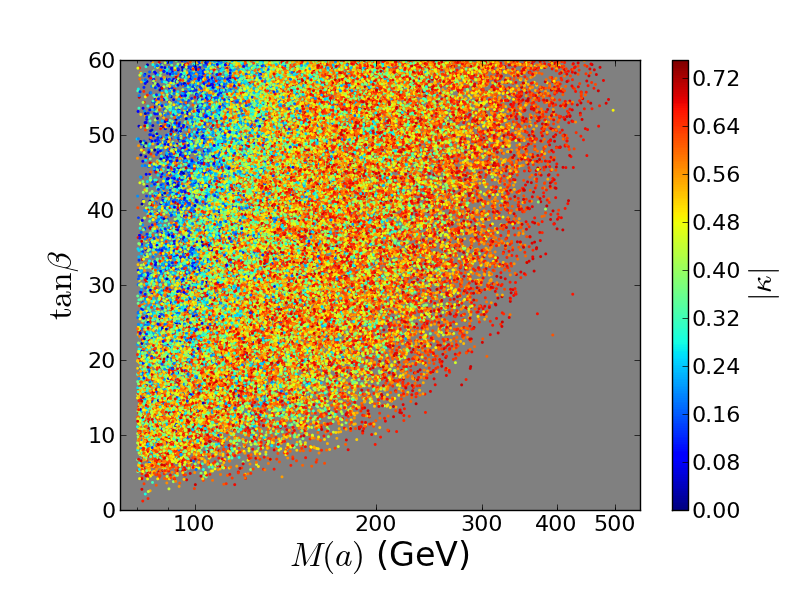}}
\vspace*{-0.10cm}
\caption{Values of the physical quantities associated with the $2\times 2$ mixing matrix of the CP-odd Higgs sector, depicting the 
allowed regions that describe the FGCE.   Top-left panel: $\sin\theta_a - m_a$ plane, color-coded by the value of $|\kappa|$. 
Top right: $\kappa - m_a$ plane, color-coded by the value of $|\sin\theta_a|$.
Bottom left: $\tan\beta - m_a$ plane, color-coded by the value of $|\sin\theta_a|$.  Bottom right: $\tan\beta-m_a$ plane,
color-coded by the value of $|\kappa|$.}
\label{fig7}
\end{figure}

\begin{figure}[htbp]
\centerline{\includegraphics[width=3.5in]{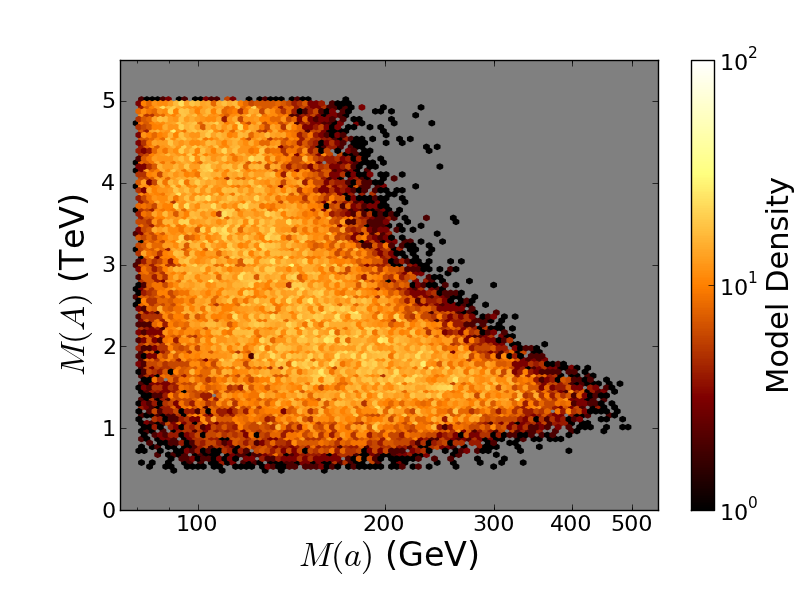}
\hspace{0.30cm}
\includegraphics[width=3.5in]{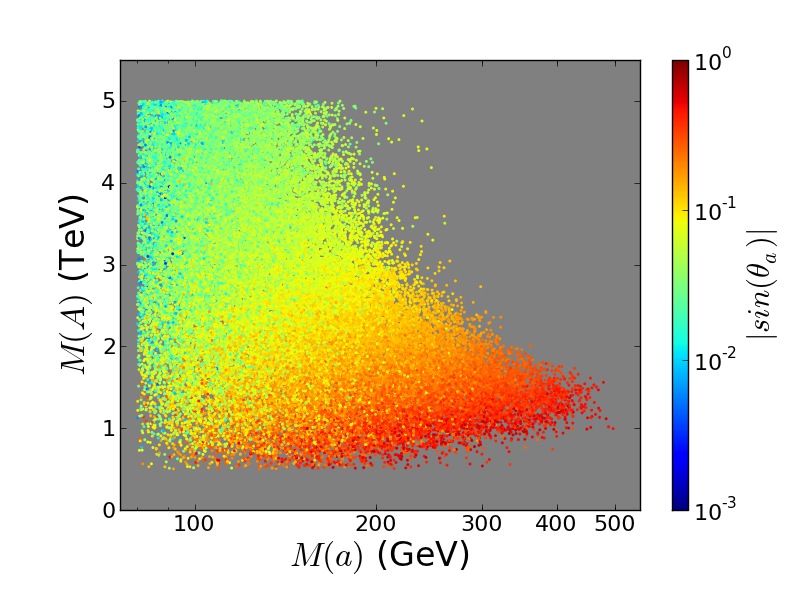}}
\vspace*{-0.10cm}
\caption{Values of the physical quantities associated with the $2\times 2$ mixing matrix of the CP-odd Higgs sector, depicting the 
allowed regions for models describing the FGCE.   Left panel: $m_A - m_a$ plane, color-coded by the density of allowed models in each
bin. Right panel: $m_A - m_a$ plane, color-coded by the value of $|\sin\theta_a|$.}
\label{fig8}
\end{figure}

We now investigate how the restricted ranges of the fundamental parameters influence the values of the physical quantities of interest. Phenomenologically,
one of the most important aspects of our general NMSSM scenario is the mixing within the 
CP-odd and CP-even Higgs sectors; this is described by the angle $\sin \theta_a$ as well as the closely related 
quantity $\sin \theta_h$, which essentially measures the Higgs 
singlet content of the $\sim 125$ GeV Higgs boson, $h$. In particular, we are interested in the correlation of $\sin \theta_a$ 
with the masses of the CP-odd scalars $a,~A$. Fig.~\ref{fig7} 
and the last panel of Fig.~\ref{fig5} show these results in some detail and there are several important observations to make: 
($i$) The light pseudoscalar mass can range up to $\sim$ 500 GeV, although the available parameter space becomes increasingly restricted as the pseudoscalar mass increases. We will examine this point in more detail below.  ($ii$) Although we 
expect that $|\sin \theta_a| \sim 0.1$ based on the simpler analysis above, we see that values as small as $\sim 0.001$ are possible, as are values as large as $\sim 0.7$ in rare cases. Of 
course the value of the mixing is highly correlated with $m_a$ as can be seen in the top panels of Fig.~\ref{fig7}. 
For fixed values of $\kappa$, as $m_a$ decreases we see that
smaller values of $|\sin \theta_a|$ are 
necessary to obtain the observed relic density. Similarly, the larger the value of $|\kappa|$ 
for fixed $m_a$ the smaller $|\sin \theta_a|$ is allowed 
to be.  ($iii$) Furthermore, the region with large $\tan \beta$ and low $m_a$ allows for both $|\kappa|$ and $|\sin\theta_a|$ to be relatively small, 
whereas larger values of $\tan \beta$ and $|\kappa|$ permit a large $m_a$. This is not 
too surprising as the dark matter annihilation amplitude scales roughly as $\sim \kappa \tan \beta \sin \theta_a /(m_a^2-2m_\chi^2)$ 
away from the $a$ pole.  Since models with $m_a$ near its upper limit require nearly maximal values of $|\sin \theta_a|$, $|\kappa|$, and $\tan \beta$, it is not possible to increase the pseudoscalar mass significantly beyond this limit while maintaining coupling perturbativity.  ($iv$) The last panel in Fig.~\ref{fig5} 
shows that $\sin \theta_a$ is also correlated with the values of $\lambda$ and $A_\lambda$; this is reasonable since 
in the limit that the LSP mass can be neglected, the off-diagonal 
entry in the effective $2\times 2$ CP-odd Higgs matrix is $\lambda A_\lambda v$ as can be seen above. ($v$) 
Of course, {\it both} eigenvalues of the CP-odd Higgs mass matrix 
are correlated with $\sin \theta_a$. First, we see in the left panel of Fig.~\ref{fig8} that the 
region with the highest model density in the $m_a-m_A$ plane occurs at 
relatively small $m_a \sim 100$ GeV, and very large values of $m_A$ of order a few TeV. The second panel in 
Fig.~\ref{fig8} displays the values of $|\sin \theta_a|$ in the $m_{a,A}$ 
plane. Obviously for fixed $m_A$, as $m_a$ increases $|\sin \theta_a|$ must increase accordingly in order to satisfy the relic density 
measurement. On the other hand, increasing $m_A$ with $m_a$ fixed lowers 
the value of $|\sin \theta_a|$ as we would expect from the diagonalization of a $2\times 2$ real symmetric matrix. The lower bound on $M_A$ from LHC searches increases with $m_a$, since heavy pseudoscalar masses require larger values of $\tan \beta$ (for which the LHC searches are most effective), as we saw in Fig.~\ref{fig7}. As a result, the heavy pseudoscalar mass becomes increasingly constrained as $m_a$ increases. 

\begin{figure}[htbp]
\centerline{\includegraphics[width=3.5in]{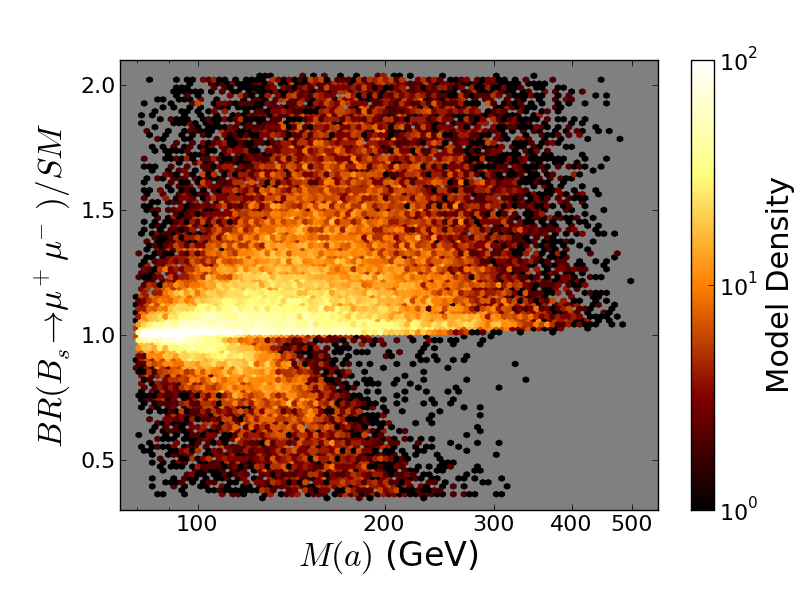}
\hspace{0.30cm}
\includegraphics[width=3.5in]{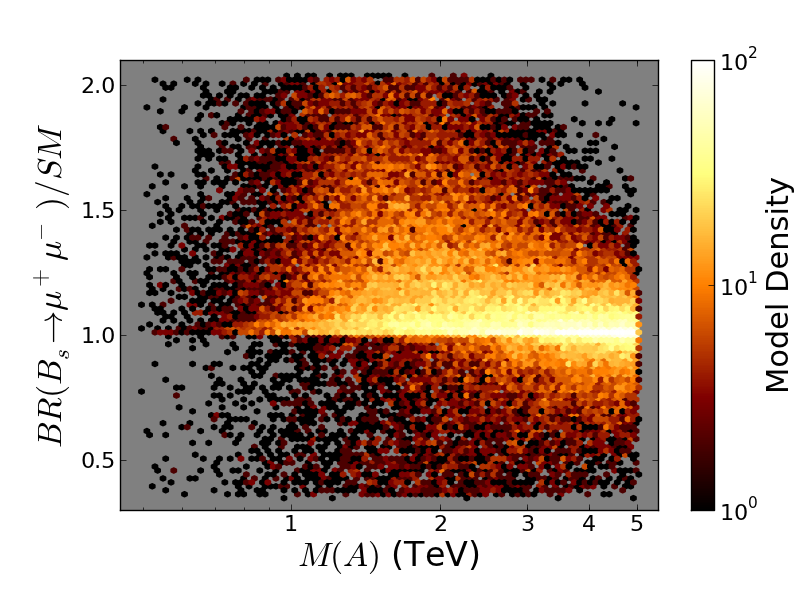}}
\vspace*{0.50cm}
\centerline{\includegraphics[width=3.5in]{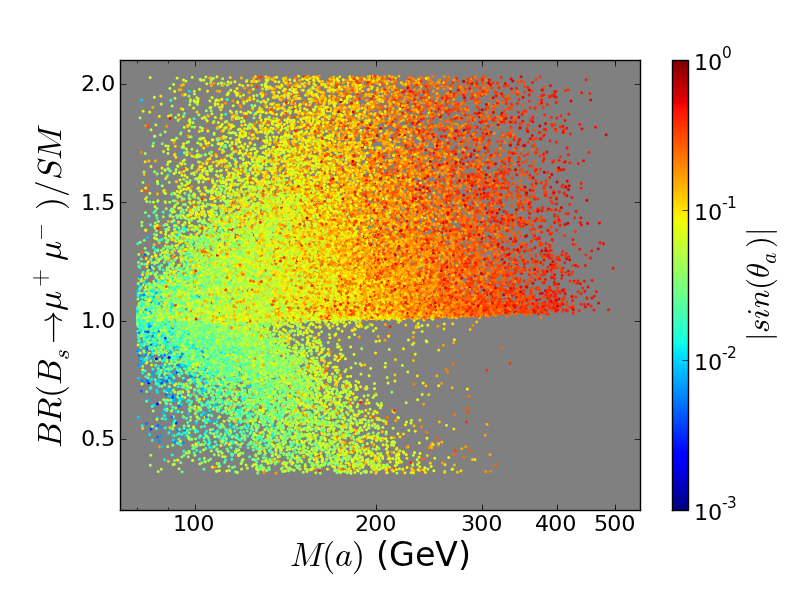}}
\vspace*{-0.10cm}
\caption{Branching fraction for the decay $B_s\to\mu^+\mu^-$ as a function of the pseudoscalar mass $m_a$ ($m_A$) in the top-left (right) panel,
color-coded by the density of allowed models in each bin.  The bottom panel shows the same axis as the upper-left panel, but now color-coded
by the value of $|\sin\theta_a|$. In each case we normalize the rate to its SM value as computed by {\tt NMSSMTools}.}
\label{fig9}
\end{figure}

The observed value of the branching fraction $BR(B_s\to \mu^+ \mu^-$) is 
quite close to the SM prediction and provides a significant 
constraint on the NMSSM parameter space, particularly for the CP-odd Higgs sector. When this sector supplies the 
dominant SUSY contribution to this decay the amplitude behaves 
as $\sim \tan^3 \beta ~(\sin^2 \theta_a/m_a^2 + \cos^2 \theta_a/M_A^2)$ and the MSSM limit is recovered when 
$\theta_a \to 0$. The first two panels in Fig.~\ref{fig9} present the value of $BR(B_s\to \mu^+ \mu^-)$ as a function of both
pseudoscalar masses, $m_{a,A}$, taken as a ratio to the SM value and color-coded according to the density of allowed models
in each bin.  Here we see that the bulk of our models predict essentially 
the same BF as does the SM, but we do have tails stretching both below and above this value. Note that as $m_A$ increases 
the distribution of predicted branching fractions narrows appreciably and becomes more localized about the SM expectation, 
while for large $m_a$ there is a bifurcation in the set of predictions producing a gap below the SM expectation.
The deviation from the SM prediction depends strongly on $\mu_{eff}$, with positive values of $\mu_{eff}$ enhancing the decay rate and negative values suppressing it compared withe the SM expectation, probably as a result of interference with the SM amplitude.  The lower panel in Fig.~\ref{fig9} shows the behavior of $BR(B_s\to \mu^+ \mu^-)$ with
respect to the value of the mixing $\sin\theta_a$.  As expected, we see that as $m_a$ increases
larger values of $|\sin \theta_a|$ are allowed, 
and that for a fixed value of $m_a$ larger values of $|\sin \theta_a|$ generally increase the branching fraction. 

\begin{figure}[htbp]
\centerline{\includegraphics[width=3.5in]{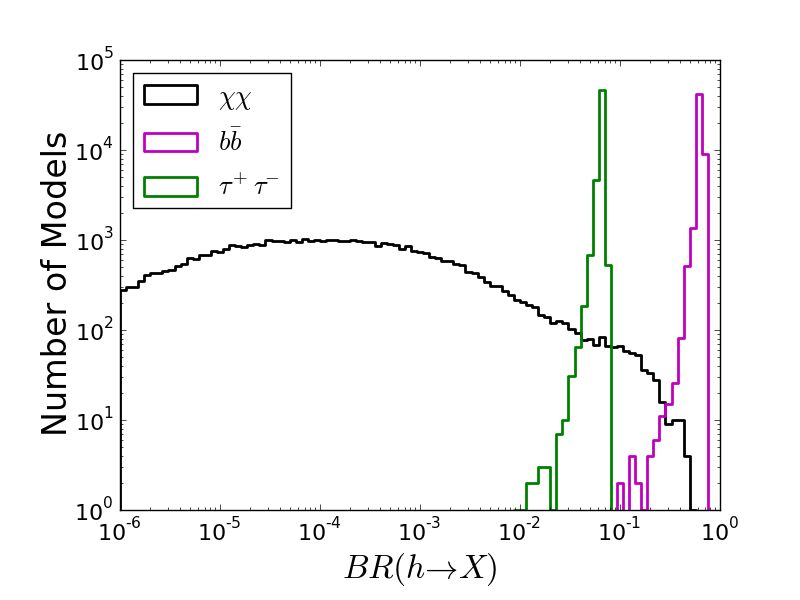}
\hspace{0.30cm}
\includegraphics[width=3.5in]{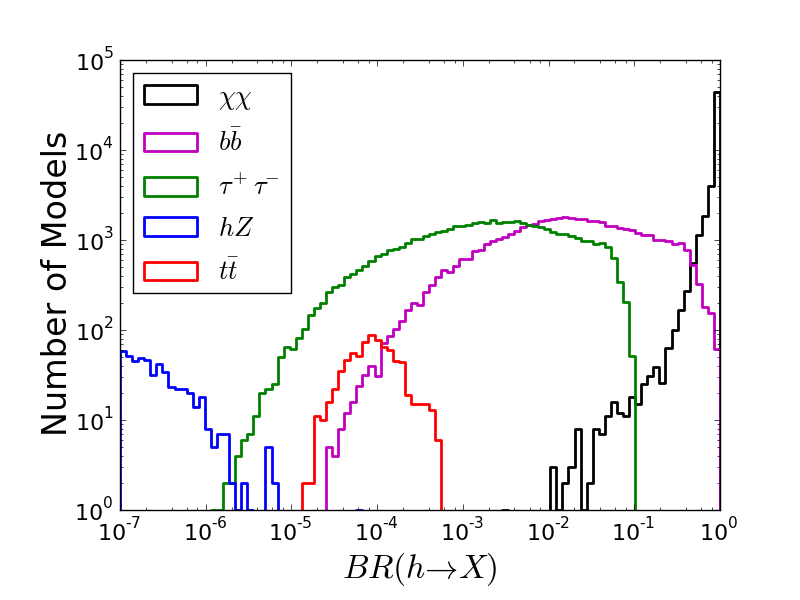}}
\vspace*{0.50cm}
\centerline{\includegraphics[width=3.5in]{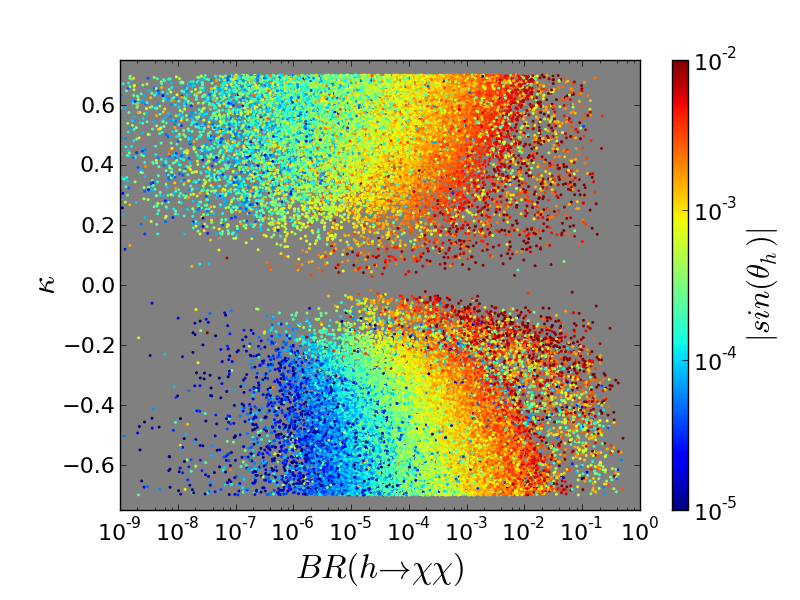}}
\vspace*{-0.10cm}
\caption{Histograms of the branching fractions of $h$ (top-left panel) and $a$ (top-right panel) to various SM and SUSY final states in our NMSSM model set.  The bottom panel presents the dependence of $BR(h\to\chi\chi)$ on $\kappa$ and the scalar singlet-doublet mixing $|\sin\theta_h|$.}
\label{fig10}
\end{figure}

Mixing between the doublet and singlet fields within both the CP-odd and CP-even Higgs sectors influences the 
decay properties of the lighter states $a$ and $h$. The top two panels in 
Fig.~\ref{fig10} display histograms of the branching fractions for the decays $h,a \to b\bar b, \tau^+\tau^-$ and LSP pairs $\chi\chi$, along with the $a \to t\bar t, Zh$ decays when kinematically allowed.
In the case of the light CP-even Higgs, we 
see that the branching fractions for both the $b\bar b$ and $\tau^+\tau^-$ final states are generally very close to their SM values since 
$|\sin \theta_h|$ is generally very small, below $\leq 10^{-2}$.  When 
this mixing takes on its larger values, the 125 GeV Higgs picks up a significant singlet component which is sufficient to allow for a
sizable branching fraction into LSP pairs.  However, for most models the $BR(h\to\chi\chi)$ usually lies below 
$1\%$ and is thus likely to be inaccessible, even to measurements made at the ILC\cite{snowhiggs}.  Models with larger branching fractions to the $\chi\chi$ final state have correspondingly reduced branching fractions to $b\bar b$ and $\tau^+\tau^-$, resulting in the tails in the corresponding distributions as observed in the top-left 
panel of Fig.~\ref{fig10}. The bottom panel in this Figure shows the dependence of $BR(h\to\chi\chi)$ on the value of $|\kappa|$
and the mixing $|\sin\theta_h|$.  We see that
increasing the magnitude of $\kappa$, which controls the strength of the singlet-LSP coupling, influences the branching fraction of
$h$  into LSP pairs; in particular, for a fixed range of values 
of $|\sin \theta_h|$, larger values of $|\kappa|$ result in a corresponding increase in the $h$ branching fraction to LSP pairs. For the 
light pseudoscalar $a$, 
the reverse situation holds - when $|\sin \theta_a|$ is small, $a$ is mostly singlet and nearly always decays to LSP pairs as seen in the top-right panel of Fig.~\ref{fig10}. 
However, as observed above $|\sin \theta_a|$ can 
be quite large, $\gtrsim 0.1-0.5$, in which case the $a$ can have significant branching fractions to the $b\bar b$ and $\tau^+\tau^-$ final states, although the branching fractions to $Zh$ and $t\bar{t}$ remain small. 
For most models, however, even the $b\bar{b}$ and $\tau^+\tau^-$ branching fractions are quite small, with, \eg, $BR(a\to\tau^+\tau^-)$ typically below $\sim 1\%$.  

\begin{figure}[htbp]
\centerline{\includegraphics[width=3.5in]{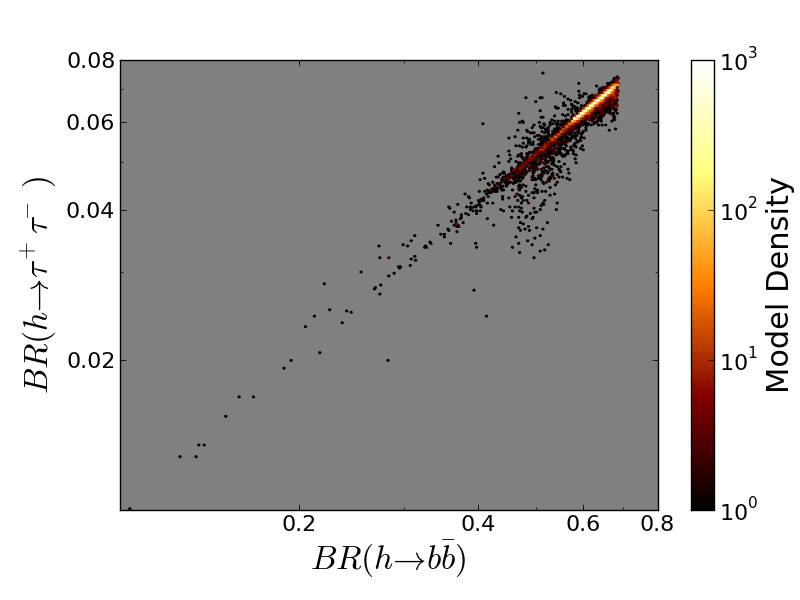}
\hspace{0.30cm}
\includegraphics[width=3.5in]{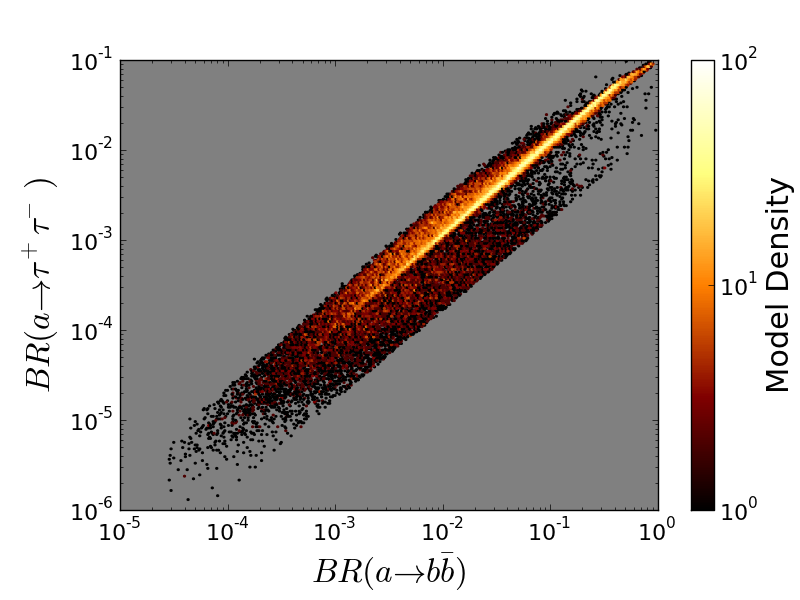}}
\vspace*{0.50cm}
\centerline{\includegraphics[width=3.5in]{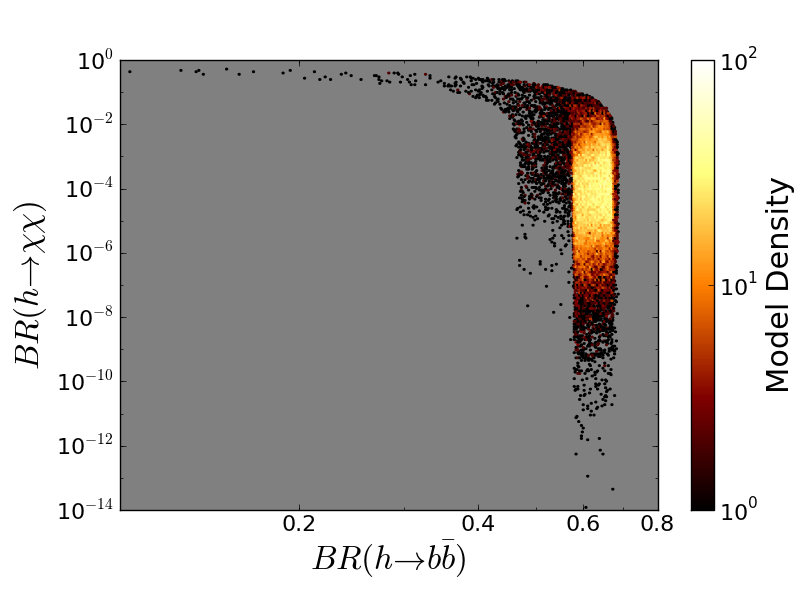}
\hspace{0.30cm}
\includegraphics[width=3.5in]{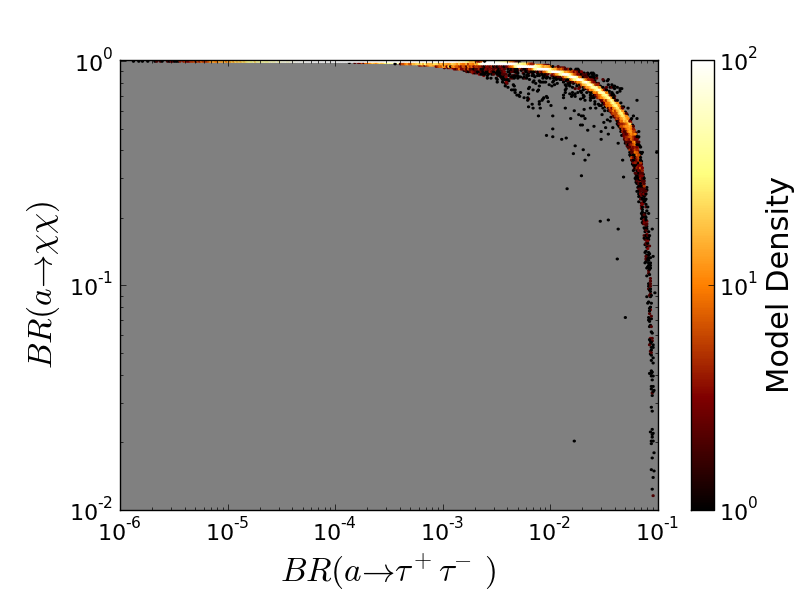}}
\vspace*{-0.10cm}
\caption{Correlations of branching fractions for the $h$ and $a$ into various final states, color-coded by the density of 
allowed models in each bin. Top-left panel: $BR(h\to\tau^+\tau^-)$ as a function of $BR(h\to b\bar b)$.  Top-right panel:
$BR(a\to\tau^+\tau^-)$ as a function of $BR(a\to b\bar b)$.  Bottom-left panel: $BR(h\to\chi\chi)$ as a function of $BR(h\to b\bar b)$.
Bottom-right panel: $BR(a\to\chi\chi)$ as a function of $BR(a\to \tau^+\tau^-)$.}
\label{fig11}
\end{figure}

Figure~\ref{fig11} shows the correlations among the $h$ and $a$ branching fractions into the $b\bar b$, $\tau^+\tau^-$ and $\chi\chi$ final states; 
these results exhibit the behavior discussed above.
We see that the branching fractions for the $b\bar b$ and $\tau^+\tau^-$ modes are generally highly correlated for both $h$ 
and $a$, although there are a few exceptions due 
to the presence of large radiative corrections. The lower two panels of this Figure show the anti-correlation of 
both the $h$ and $a$ decays to LSP pairs with their 
decays to SM final states. Here we see, \eg, that the $h \to b\bar{b}$ branching fraction remains in the SM range until the singlet admixture  
becomes large enough to generate a sizable decay rate into LSP pairs. 
Note that $BR(a\to\chi\chi)$ is nearly unity for most of the models in this set and thus the decay rates to SM final
states are generally very small.

\begin{figure}[htbp]
\centerline{\includegraphics[width=3.5in]{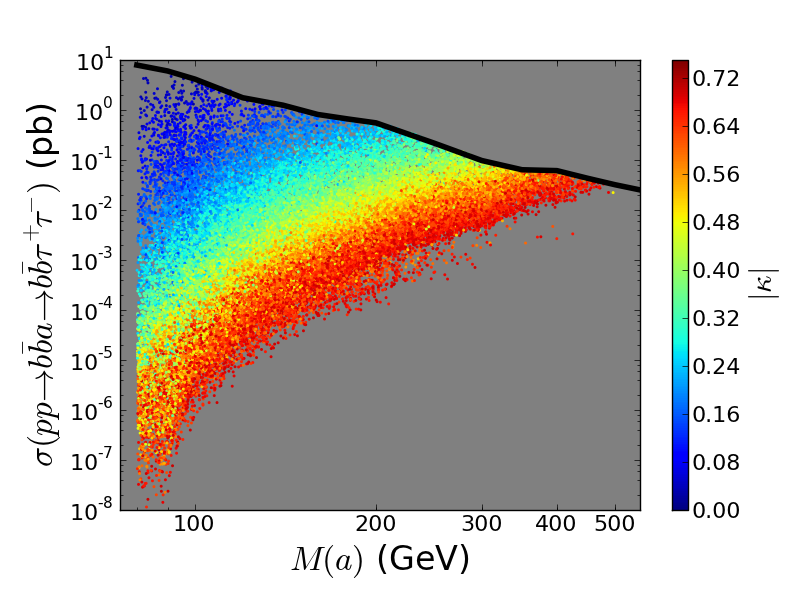}
\hspace{0.30cm}
\includegraphics[width=3.5in]{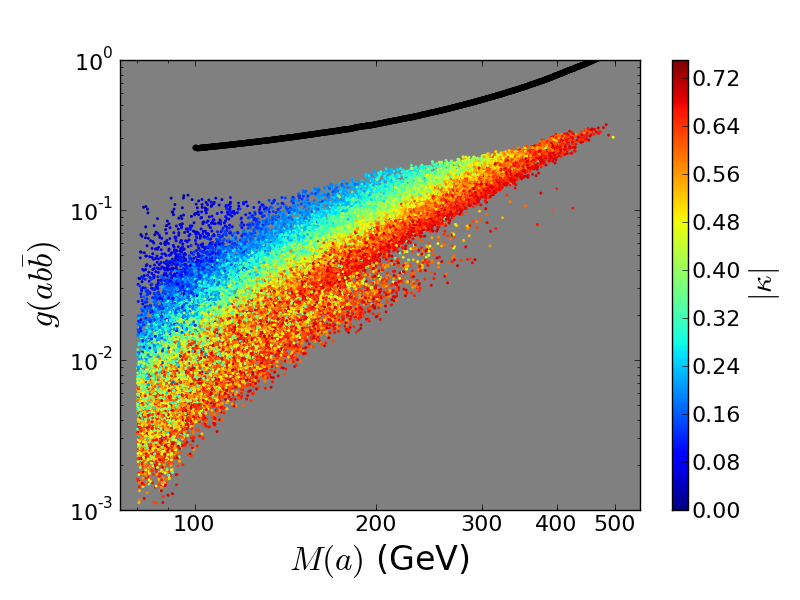}}
\vspace*{-0.10cm}
\caption{Left panel: The signal strength for $pp\to b\bar{b} a$, with $a$ decaying to $\tau^+\tau^-$, as a function of $m_a$. The CMS limit on this process~\cite{cmstautau} is shown as a solid black line. Right panel: The $ab\bar b$ coupling as a function of $m_a$. A limit on the $ab\bar b$ coupling, assuming a $\sim 100\%$ BF to LSP pairs, was obtained by rescaling the limit in Figure 5 of~\cite{eder}, and is shown as the solid black line. Both panels are color-coded by the value of $|\kappa|$}
\label{fig12}
\end{figure}

Now that we understand the decays of the light pseudoscalar $a$, we can examine
the potential constraints on our NMSSM model set arising from 
searches for additional Higgs fields at the LHC. We note that {\tt NMSSMTools} 
implements the well-known constraints from searches for the MSSM states $H$ and $A$ decaying to $\tau^+ \tau^-$, in both the gluon fusion and $b\bar{b}$-associated production channels, so that the current limits were applied during the model generation process. In Fig.~\ref{fig12} we examine constraints on $b\bar{b}$-associated production of $a$ and their dependence on $|\kappa|$. The first panel of the Figure shows the rate for associated production with $a$ decaying to $\tau^+ \tau^-$. We see the rate decreases dramatically with increasing $|\kappa|$. This rate decrease results from the corresponding increase in the three-point coupling $a \chi \chi$, which has three important effects. First of all, a large $a \chi \chi$ coupling reduces the $a$ branching fraction to $\tau^+ \tau^-$ for a given $a \tau^+ \tau^-$ coupling. Second, models with a large $a \chi \chi$ coupling need a smaller value of the $a b \bar{b}$ coupling to obtain the correct annihilation cross section, resulting in a diminished $b\bar{b}$-associated production cross section. Third, since both the $a b \bar{b}$ and $a \tau^+ \tau^-$ couplings are proportional to $\tan \beta \times \sin\theta_a$, reducing the $a b \bar{b}$ coupling also reduces the $a \tau^+ \tau^-$ coupling, further suppressing the branching fraction to $\tau^+ \tau^-$. We see that the LHC search provides an important constraint on models with small values of $|\kappa|$, especially at larger values of $m_a$. We do not show the constraints from production of $a$ via gluon fusion since they are generally weaker than the constraints from $b\bar{b}$-associated production.

Of course, if $a$ decays mostly into LSP pairs then $b\bar b a$ production
may be observable as $b\bar b+$MET; this channel has been recently discussed in Ref.~\cite{eder}.
In the limit where the $a \chi \chi$ coupling is much larger than the $ab\bar b$ coupling (so that $BR(a \to \chi \chi)\simeq 1$), which is valid for most models in our sample, the signal strength in this channel is completely described by the $a$ mass and the $ab\bar b$ coupling. With this assumption, we can compare the $ab\bar b$ coupling in our models to the limit in the lower panel of Figure 5 in \cite{eder} with appropriate rescaling. We show the result, including its $|\kappa|$ dependence, in the right panel of Fig.~\ref{fig12}. Once again we see that models with large values of $|\kappa|$ are more difficult to observe, however the dependence is much more mild since the branching fraction for $a$ to LSP pairs is typically $\sim 1$ and therefore essentially independent of $|\kappa|$, meaning that the only impact of $|\kappa|$ is through its effect on the $a b \bar{b}$ coupling.
We see that our models are safely below the bound, although searches specifically targeted at this signal could conceivably be competitive with the ditau searches in the future.

\begin{figure}[htbp]
\centerline{\includegraphics[width=3.5in]{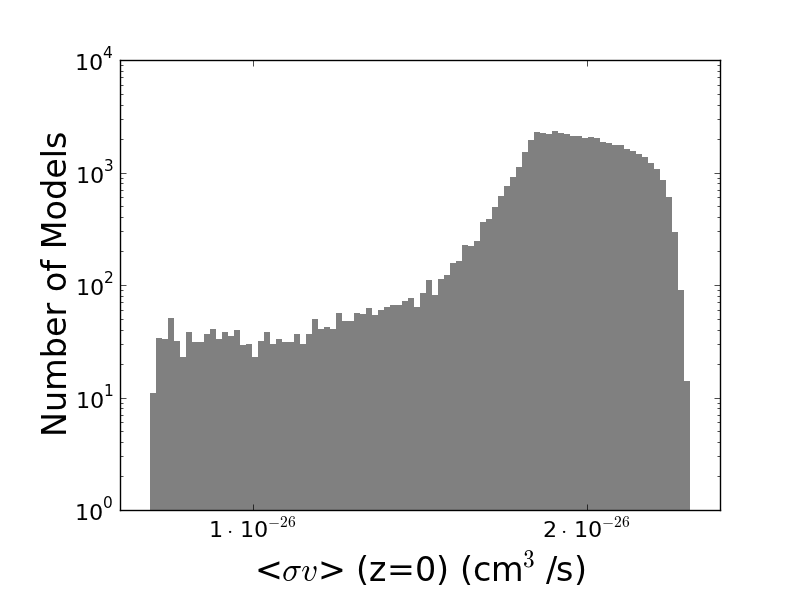}
\hspace{0.30cm}
\includegraphics[width=3.5in]{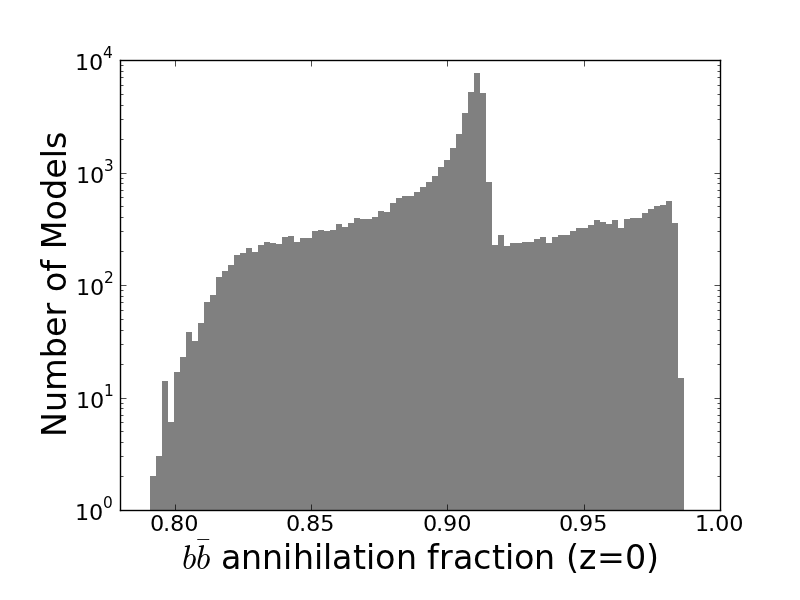}}
\vspace*{0.50cm}
\centerline{\includegraphics[width=3.5in]{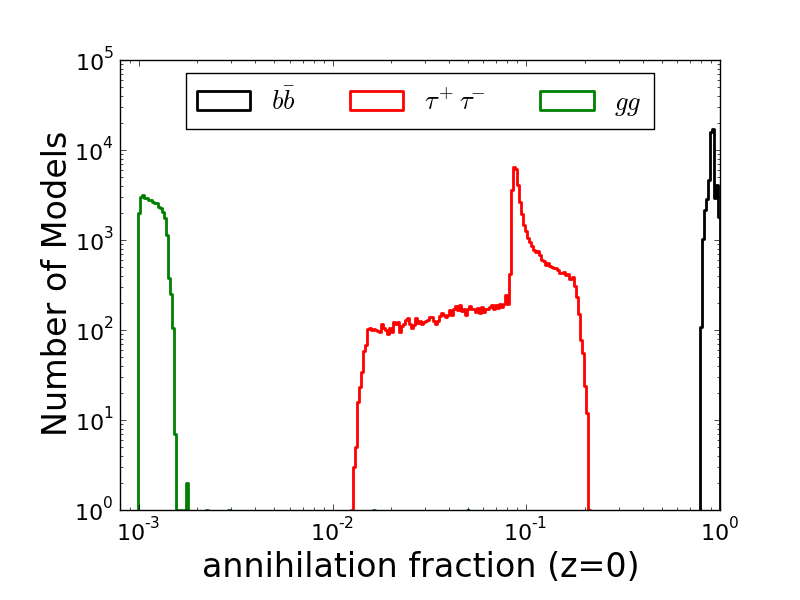}
\hspace{0.30cm}
\includegraphics[width=3.5in]{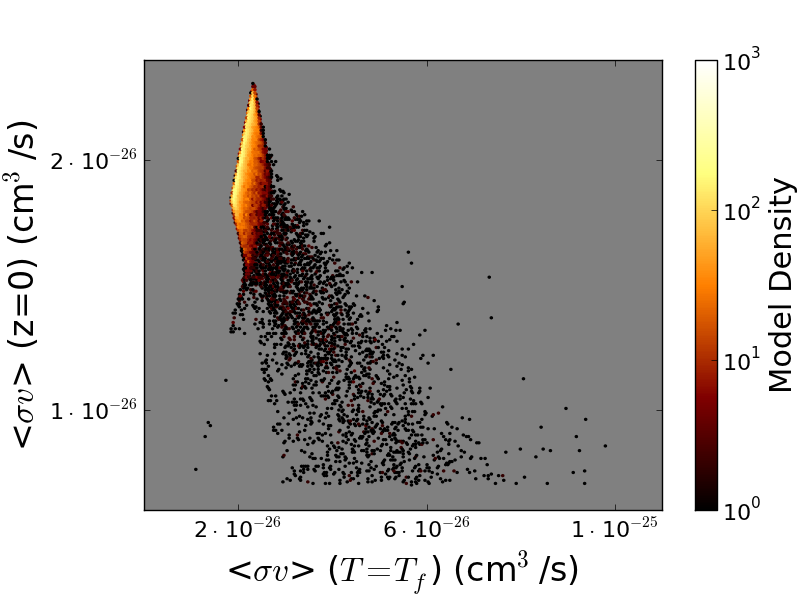}}
\vspace*{-0.10cm}
\caption{Predicted dark matter annihilation rates for our NMSSM model points.  Top-left panel:  Histogram of the inclusive present-day annihilation cross section.  Top-right panel: Histogram of the present-day annihilation rate in the $b\bar b$ final states, taken as a ratio to the inclusive annihilation rate.  Bottom-left panel: Histograms of the present-day annihilation rate into $b\bar b$, $\tau^+\tau^-$,
and $gg$ final states, taken as a ratio to the inclusive annihilation rate.  Bottom-right panel: Correlation between the inclusive present-day annihilation rate and
its value at freeze-out.}
\label{fig13}
\end{figure}

We now study the properties of the dark matter sector for our NMSSM model points.
The top-left panel in Fig.~\ref{fig13} shows the distribution for the present-day total LSP pair annihilation cross 
section for our NMSSM model set. Recall that the lower limit of this histogram was set by the requirement 
that we reproduce the observed FGCE based on the analysis of Ref.~\cite{recentgce}. Note that the annihilation rate for the
majority of the models lies 
above $<\!\sigma v\!> = 1.5\times 10^{-26}$~cm$^3$s$^{-1}$. The top-right panel presents a histogram of the annihilation fraction to the $b\bar b$ final state.  By design, the most 
common final state for the LSP annihilation process is indeed $b\bar b$, which typically comprises $\sim 90\%$ of the total rate. However, as 
shown in the bottom-left panel of Fig.~\ref{fig13}, other final states, such as $\tau^+\tau^-$ and $gg$, can also have significant rates, with the $\tau^+\tau^-$
channel accounting for $\sim 10\%$ of the annihilation rate in many models.
The bottom-right panel shows the correlation between the present-day dark matter annihilation cross section 
and that at freeze-out, which accounts for the observed relic density (recall that we employed the 
{\tt NMSSMTools} version of {\tt micrOMEGAs}). We see that the bulk of the models reside in the diamond-shaped region in the 
upper left corner of the panel, where the two annihilation cross sections have comparable values with the present-day rate being
slightly smaller as expected. However, we also find a long tail of model points with larger freeze-out cross sections and smaller 
present-day annihilation rates. In these cases, we find that 
$m_a$ is close enough to $2m_\chi$ for the annihilation rate to benefit significantly from the $a$ pole during freeze-out, resulting in a suppression of the present-day annihilation rate which is not resonantly enhanced.  Some of these models also have a slightly elevated Higgsino 
content of the LSP, so that $Z$ exchange can also play a small role during the annihilation process in the early universe. 

\begin{figure}[htbp]
\centerline{\includegraphics[width=3.5in]{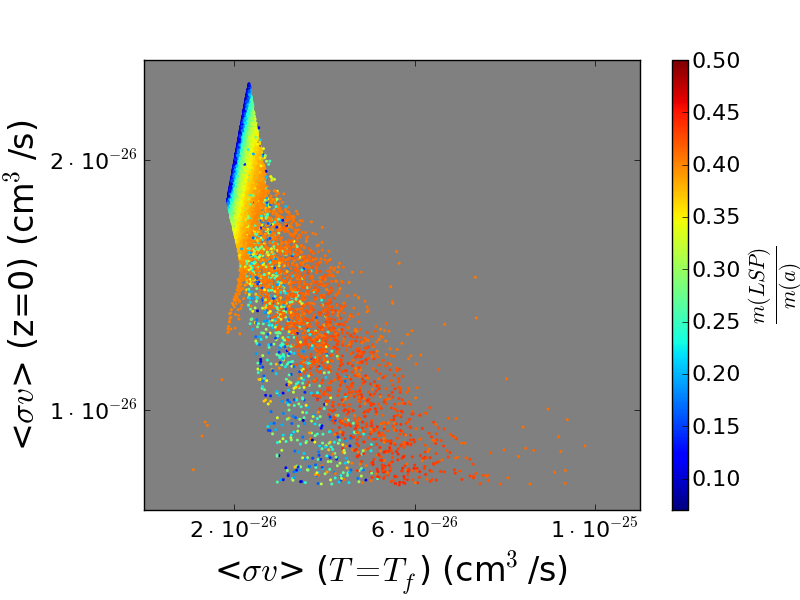}
\hspace{0.30cm}
\includegraphics[width=3.5in]{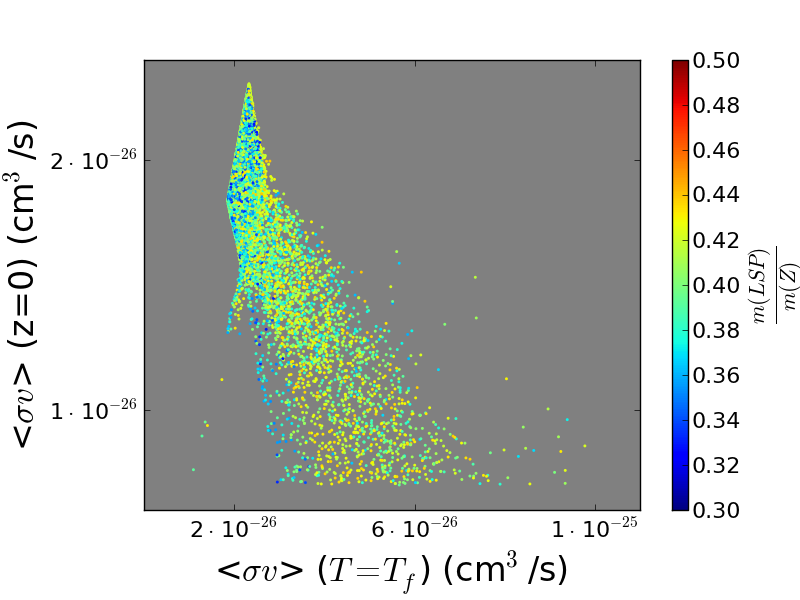}}
\vspace*{-0.10cm}
\caption{Correlation between the inclusive present-day annihilation rate and its value during freeze-out, color-coded by the ratio of the LSP mass to the mass of the $a$ (left) and $Z$ (right) bosons.}
\label{fig14}
\end{figure}

In Fig.~\ref{fig14} we further examine the correlation between the present-day and freeze-out dark matter
annihilation cross sections.  Here, the models are colored-coded by the values of the LSP mass relative to both the $a$ and the $Z$, 
so that we can ascertain the effects of these specific resonance exchanges. 
As expected, we see in the left 
panel that the tail of the distribution that descends to lower present-day annihilation rates and larger rates during freeze-out occurs as
$2m_\chi$ approaches $m_a$.  This indicates that the $a$ pole is indeed enhancing the early universe annihilation rate for these models as we expected. 
Interestingly, at the bottom of this long tail we find a set of models (colored blue and green) where 
the LSP is quite light relative to the $a$, but where the present-day cross section is still suppressed.  The right panel of Fig.~\ref{fig14} displays the same 
two annihilation cross sections with the points now color-coded by the ratio of the LSP and $Z$ masses. Here, the range of mass ratios is {\it a priori} 
narrow as the LSP was required to lie in the 30-40 GeV mass region, however some variation 
is noticeable with models having lighter LSPs tending to cluster on the top-left side of the distribution. In this panel
we see that most of the models in the tail have masses not too far from $\sim 0.4 m_Z$ so that the influence of the $Z$ 
pole is likely non-negligible, accounting for the models which are far from the $a$ pole but still have suppressed present-day cross sections.

\begin{figure}[htbp]
\centerline{\includegraphics[width=3.5in]{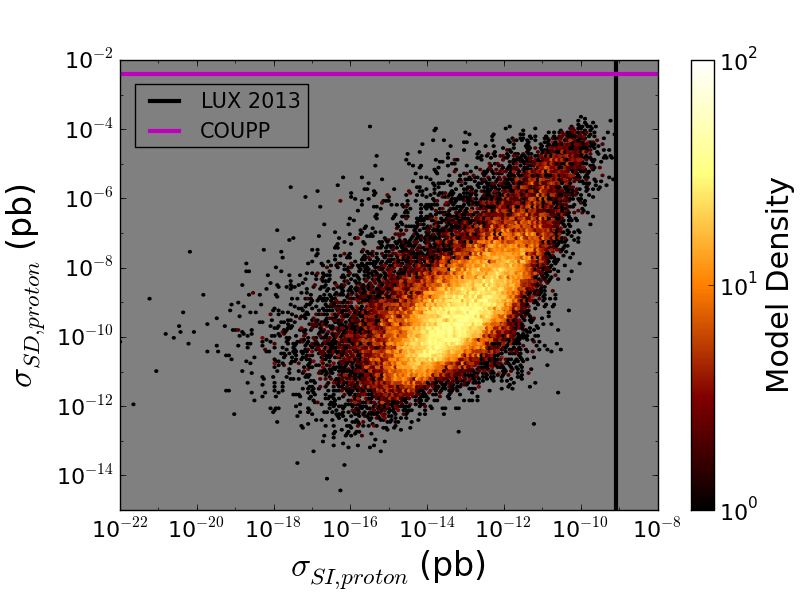}
\hspace{0.30cm}
\includegraphics[width=3.5in]{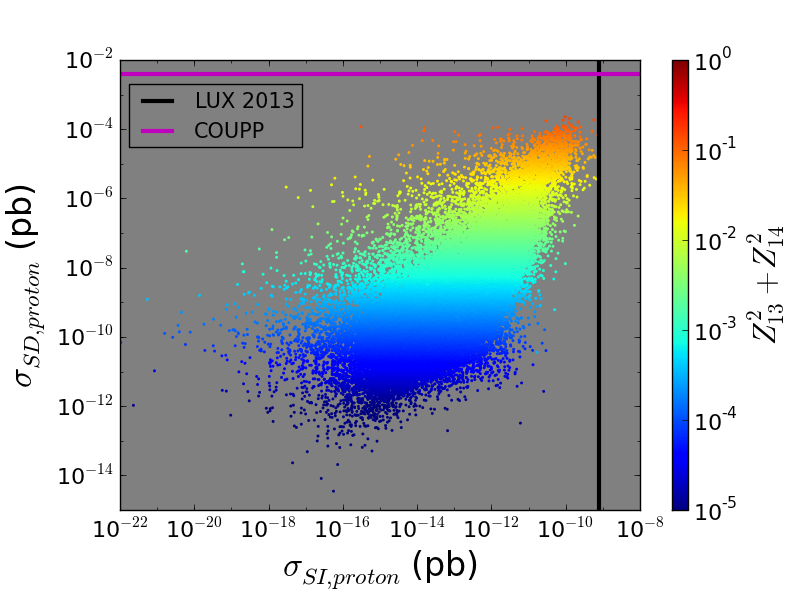}}
\vspace*{0.50cm}
\centerline{\includegraphics[width=3.5in]{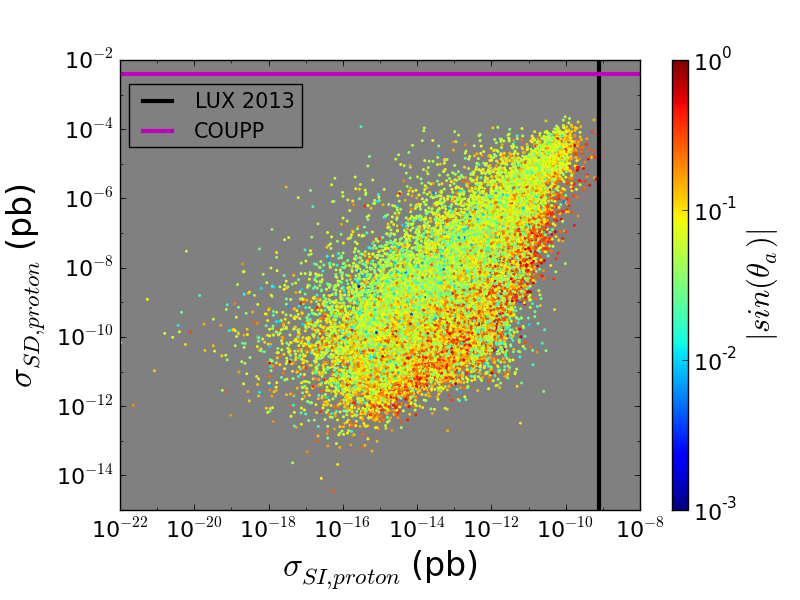}
\hspace{0.30cm}
\includegraphics[width=3.5in]{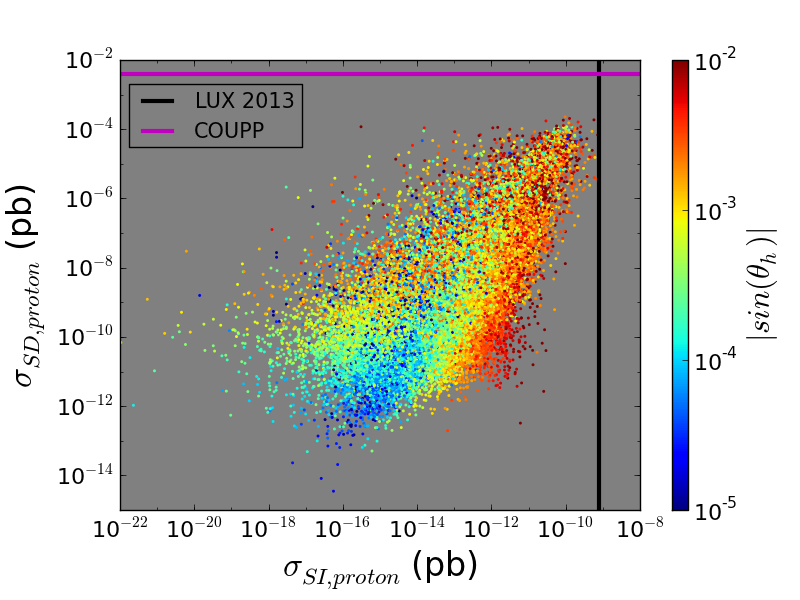}}
\vspace*{-0.10cm}
\caption{Scatter plots of the spin-dependent vs spin-independent scattering cross sections for our NMSSM model points, color-coded by the density of allowed models in each bin (top left), the Higgsino content of the LSP,  $(|Z_{13}|^2+|Z_{14}|^2)$ (top right), the pseudoscalar singlet-doublet mixing $|\sin\theta_a|$ (bottom left), and the scalar singlet-doublet mixing $|\sin\theta_h|$ (bottom right). Current limits from COUPP~\cite{Behnke:2012ys} and LUX~\cite{Akerib:2013tjd} are shown as horizontal and vertical lines, respectively.}
\label{fig15}
\end{figure}

Lastly, we consider both the spin-independent (SI) and spin-dependent (SD) direct detection cross sections for our model points in 
Fig.~\ref{fig15}, along with the current LUX and COUPP constraints.
The SI cross section is a measure of the interaction between the LSP and the nucleus, and in this scenario is mostly due to CP-even 
Higgs exchange resulting from mixing via, \eg, a  
non-zero value for $\sin \theta_h$ or a small Higgsino admixture in the LSP, as well as through a possible 
loop-induced $h\chi\chi$ coupling. Note that the $t$-channel exchange contribution is small since
the first and second generation squarks are quite heavy in the cases we consider here.
The SD interaction, on the other hand, can arise from $Z$ exchange via the Higgsino content of the LSP and/or through the 
exchange of the CP-odd Higgs fields, although the latter process is suppressed by the values of both the mixing angles well as $v^2$. We thus expect that both 
the SI and SD direct detection cross sections for the singlino-like LSPs in our model set will be quite small. Indeed, this is what we observe 
in the top-left panel of Fig.~\ref{fig15}, which shows the model density distribution in the SI-SD cross section plane. Here we see 
that typical values for the SI (SD) cross sections are of order $\sim 10^{-13}(10^{-10})$ pb, both of which are several orders of magnitude away 
from the current experimental bounds. Unfortunately only a small fraction of our model points will be accessible to future direct detection 
searches. The top-right panel in this Figure shows
how the magnitude of the SD cross section is strongly correlated with the Higgsino content of the 
LSP, denoted here as $Z_{13}^2+Z_{14}^2$ and represented by the point coloration.  Models where the LSP has a larger Higgsino 
content have a larger coupling to the $Z$ boson, resulting in sizable SD scattering through $Z$ exchange. On the other hand, we find that the SI cross section is apparently not sensitive to 
the Higgsino content of the LSP for this set of models.  The bottom-left panel shows the dependence of the SI and SD cross sections on the 
value of $|\sin \theta_a|$, which could be significant if $a,A$ exchange are the dominant contribution. Here we see that, at most, 
there is a modest sensitivity implying that the SD cross section is indeed dominated by $Z$ exchange. The bottom-right panel 
shows the dependence of the SI cross section on the value of $|\sin \theta_h|$. The strong dependence of the SI cross section on $|\sin \theta_h|$ shows that the SI scattering proceeds mainly through $h$ exchange, via the mixing between $h$ and the scalar singlet Higgs. However, since there are, in fact, three CP-even Higgs bosons of various masses and couplings 
contributing simultaneously to the SI cross section, certain values of their masses and mixings can result in interference between the various contributions. The effect of this interference can be seen as the orange model 
points in the top left side of the Figure, for which relatively large values of $|\sin \theta_h|$ still result in rather small SI cross sections.  

\section{Conclusions}

In this paper we have considered a set of SUSY scenarios, searching for a model that might lead to a consistent explanation of the FGCE while 
also satisfying other known experimental and theoretical constraints. We began by exploring a variety of SUSY models 
which failed to describe this excess, all for different reasons. These attempts demonstrate the significant model 
building challenge posed by the FGCE within the SUSY context.  

In the (p)MSSM a $\sim 35$ GeV neutralino can lead to the correct relic density only if it is a bino-Higgsino 
admixture that annihilates near the $Z$ pole in the early universe.{\footnote {The possibility of achieving the observed 
relic density with a bino-like LSP exchanging a very light $\tilde \tau_R$ remains viable.}} 
At the present time, dark matter annihilation takes place far from the $Z$ pole and so the prediction for the FGCE flux 
is too small by several orders of magnitude. If the LSP were instead a Dirac gaugino, this would help to increase the 
annihilation cross section by avoiding the velocity/helicity suppression that occurs in the Majorana scenario. Of course, 
in this case the LSP would need to be a very pure bino to avoid the direct detection constraints and so the observed relic 
density could only be obtained via $t$-channel sfermion exchange, in particular, through a right-handed $\tilde \tau$ with 
mass near the LEP limit. Given the hypercharge of the right-handed sbottom, as well as the lower bound on the sbottom mass, the 
desired $b\bar b$ final state would not occur at high enough rates to account for either the observed relic density or the FGCE.  
In the case of an extended gauge symmetry, such as the GUT theory $E_6$, new fermions and 
new gauge bosons are present so that the $Z$-induced direct detection constraint can be avoided if the LSP were a Dirac 
(or even Majorana) SM singlet annihilating through a heavy new gauge boson $Z'$. Unfortunately, the LHC lower bounds on a $Z'$ mass 
lead to a highly suppressed DM annihilation cross section for either of the SM singlet DM candidates present in 
$E_6$.

We next turned our attention to the NMSSM, identifying the LSP as an almost pure singlino state.  We first investigated the
$Z_3$ version of the NMSSM (with only four new parameters beyond the MSSM)
under the assumption that the DM annihilation process that produces 
the observed relic density ($s$-channel light CP-odd Higgs exchange) is the {\it same} as that which leads to the presently 
observed FGCE.  Regrettably, we showed that this scenario also proves to 
be inadequate to the task. This is due to the large Higgsino admixture which is necessarily picked up by the LSP 
due to the structure of the neutralino mixing matrix, and the insufficient parameter freedom which prevents us from avoiding this mixing in this 
scenario. Recall that once the LSP has picked up a significant Higgsino component then the $Z$ will play an important role 
in the dark matter annihilation process and one has to fine-tune the parameters to generate the observed relic density as well as
the present-day annihilation in the galactic center.  We wished to avoid such fine-tuning, motivating our 
assumption that CP-odd Higgs exchange is solely responsible for both.  Lastly, we examined 
the general NMSSM which has nine additional superpotential plus soft-breaking parameters beyond those of the 
usual (p)MSSM and thus has far greater flexibility. To better understand the various phenomenological requirements that the 
FGCE imposes on this version of the NMSSM, we began with an examination of this model at tree level. Even in this rather 
simplified case we found that it was non-trivial to meet both the dark matter requirements as well as 
other experimental constraints. However, viable solutions were shown to exist, selecting a specific region of the parameter space. 
In order to perform a more complete and detailed analysis, we made use of a modified 
version of the {\tt NMSSMTools4.3.0} code which includes radiative corrections to the various mass spectra and corresponding couplings, 
and also provided a means to calculate the relic density as well as the dark matter direct detection and present-day pair-annihilation 
cross sections in this scenario.     

We then generated a set of $6 \cdot 10^8$ points in the general NMSSM parameter space with rather loose range 
requirements, subjecting each of them to a set of global experimental constraints (collider, flavor, astrophysical, 
\etc.) and also demanding a sufficiently large present-day dark matter annihilation cross section into $b\bar b$ in order to explain 
the FGCE.  Of this set, $\sim 52.8$k model points survived these basic requirements, demonstrating that the general NMSSM with a single dark matter annihilation mechanism provides a successful SUSY scenario 
for the FGCE. We then studied the physical properties of these surviving model points in some detail. First, we examined 
the regions of the NMSSM parameter space that led to viable models, studying in particular the role played by correlations 
among these parameters and the properties of the physical Higgs states. One of the interesting results of 
this investigation is the rather wide range of possible values for the light CP-odd Higgs mass, $m_a$, (which is correlated with the 
values of $\kappa$ and $\sin \theta_a$, and also to some extent $\tan \beta$)  allowed by the constraints. 
Since the $a$ has a large singlet component and the LSP is dominantly singlino, $a$ principally decays to LSP pairs. This makes it difficult to observe at the LHC, where searches for b-jets and missing energy face large backgrounds. Although the best constraints come from di-tau resonance searches, future searches for invisible $a$ decays could also provide some sensitivity.  In contrast, the light CP-even, $\sim 125$ GeV Higgs generally has a rather 
weak, mixing-induced coupling to the LSP, so that the branching fraction for $h\to \chi \chi$ typically lies far below $1\%$; hence $h$ has 
properties quite close to that of the SM Higgs, as is the case in the MSSM. 

Lastly, we studied the dark matter and flavor properties in our NMSSM model set.
Since the LSP only couples to the SM via mixing with the Higgsinos, 
and/or mixing amongst the Higgs fields, both the SI and SD direct detection cross sections are found to be quite small, 
\ie, several orders of magnitude away from the current experimental bounds, in this scenario. We find that only a small fraction of our NMSSM models 
would be accessible to future direct detection searches.  Rare $B$ decays, such as $B_s \to \mu^+\mu^-$, can probe this parameter space to 
some extent. Although most of the models yield branching fractions quite close to the SM prediction, there are a significant number of cases where the 
predictions differ from the SM by $\sim 50\%$. Once the theoretical and experimental uncertainties are under control, $B_s\to\mu^+\mu^-$ could
provide a signal for this subset of models.   

In summary, we find that it is non-trivial to build a UV-complete model based on SUSY that describes the FGCE.
However, we find that the general NMSSM easily accommodates the data and that the data selects a specific region of parameter space.
We find that increasing the LSP mass to 70 GeV does not affect these conclusions.
Unfortunately, the signatures elsewhere, such as in Higgs physics at the LHC and in direct detection, are small and will be difficult, if not
impossible, to detect.  It is possible that a signature could be observed in the flavor sector.

Indirect detection thus provides the best probe of this scenario.
Given that the FGCE dictates the dark matter annihilation process cross section, \Fermi~ should soon see some gamma ray excesses 
elsewhere, \eg, in the dark matter searches using Dwarf Galaxies.
We hope that further data will confirm the signature observed at the galactic center and that dark matter will soon be
discovered!

\section*{Acknowledgments}

This work was supported by the U.S. Department of Energy, Contract DE-AC02-76SF00515 and Grant ER41990.
We would like to thank Elliott Bloom, Ahmed Ismail, Tracy Slatyer, Tim Tait, and Matthew Wood for discussions.

\end{document}